\documentclass[aps,twocolumn,amsmath,amssymb,floatfix,showpacs,superscriptaddress]{revtex4}

\usepackage{bm,amsmath,amssymb,amsfonts}

\usepackage[dvips]{graphicx}
\usepackage{array}

\usepackage{color}
\definecolor{brown}{rgb}{0.70,0.30,0.05}
\usepackage{hyperref}
\hypersetup{
    pdfnewwindow=true,      
    colorlinks=true,       
    linkcolor=blue,          
    citecolor=red,        
    filecolor=blue,      
    urlcolor=blue        
}

\begin{document}

\title{Spin filter for arbitrary spins by substrate engineering}

\author{Biplab Pal}
\email{biplabpal@klyuniv.ac.in}
\affiliation{Department of Physics, University of Kalyani, Kalyani,
West Bengal-741235, India}

\author{Rudolf A. R\"{o}mer}
\email{r.roemer@warwick.ac.uk}
\affiliation{Department of Physics and Centre for Scientific Computing, 
University of Warwick, Coventry, CV4 7AL, UK}

\author{Arunava Chakrabarti}
\email{arunava_chakrabarti@yahoo.co.in}
\affiliation{Department of Physics, University of Kalyani, Kalyani,
West Bengal-741235, India}

\begin{abstract}
We design spin filters for particles with potentially arbitrary spin 
$S$ ($=1/2$, $1$, $3/2$, \ldots) using a one-dimensional periodic chain
of magnetic atoms as a quantum device. Describing the system within a 
tight-binding formalism we present an analytical method to unravel the 
analogy between a one-dimensional magnetic chain and a multi-strand 
ladder network. This analogy is crucial, and is subsequently exploited 
to engineer gaps in the energy spectrum by an appropriate choice of the 
magnetic substrate. We obtain an exact correlation between the magnitude 
of the spin of the incoming beam of particles and the magnetic moment 
of the substrate atoms in the chain desired for opening up of a spectral 
gap. Results of spin polarized transport, calculated within a transfer 
matrix formalism, are presented for particles having half-integer as 
well as higher spin states. We find that the chain can be made to act 
as a quantum device which opens a transmission window only for selected 
spin components over certain ranges of the Fermi energy, blocking them 
in the remaining part of the spectrum. The results appear to be robust 
even when the choice of the substrate atoms deviates substantially from 
the ideal situation, as verified by extending the ideas to the case of 
a `spin spiral'. Interestingly, the spin spiral geometry, apart from 
exhibiting the filtering effect, is also seen to act as a device 
flipping spins -- an effect that can be monitored by an interplay 
of the system size and the period of the spiral. Our scheme is 
applicable to ultracold quantum gases, and might inspire future 
experiments in this direction.
\end{abstract}

\pacs{72.25.Mk, 85.75.-d}

\maketitle
\section{Introduction}
\label{intro}
The idea of transporting information through electron spins instead 
of charge, \emph{spintronics}~\cite{prinz,wolf}, has opened a 
promising pathway to quantum information processing and quantum 
computation in the future. Spurred by the measurement of tunneling 
magnetoresistance in magnetic tunnel junctions~\cite{julliere,moodera}, 
and the observation of giant magneto-resistance in magnetic 
multilayers~\cite{baibich}, the search for the integration of memory 
and logic in a single storage device has taken an inspiring shape 
in the last couple of decades.

A substantial part of the existing research focuses on experiments 
related to spin polarized electron transport in nanostructures. 
The quantum confinement effect on transport of electrons was studied 
by several groups~\cite{gordon,cronenwett,yu}. Tunable spin filters 
have been developed where charge carriers with different spin states 
were separated in GaAs samples~\cite{rokhinson}. A `non-local' spin 
valve geometry was used to study spin transport in single graphene 
layers~\cite{tombros}. Experimental realizations of a quantum spin 
pump using a GaAs quantum dot~\cite{watson}, an `open' quantum dot 
driven by ac gate voltages~\cite{eduardo}, along with several other 
works such as the study of spontaneous spin polarized transport in 
magnetic nanowires~\cite{rodrigues}, or an analysis of the spin 
polarization of the linear conductance of a quantum wire spin 
filter~\cite{birkholz}, have enriched the field of spin polarized 
transport. The design of molecular wires and spin polarized 
tunneling devices~\cite{andes} is also in the cards in the current 
era of spintronics.

Needless to say, such experiments have inspired a bulk of theoretical 
investigations of spin transport, or spin polarized coherent electronic 
transport in nano structures, in model quantum dots or magnetic 
nanowires~\cite{sergueev,lu,wang,shokri1,shokri2,shokri3,maiti1,maiti2,maiti3}, or, 
in a very recent work, modeling a ferroelectric polymer grown on top of a silicene 
nanoribbon~\cite{cesar}.  
A widely adopted line of attack has been to work within a tight-binding 
formalism in which a nano wire is simulated by placing `magnetic atoms' 
in a line, and sandwiching the array between two semi-infinite magnetic 
or non-magnetic leads~\cite{shokri1,shokri2,shokri3,maiti1,maiti2,maiti3}. 
Green's function method and transfer matrix 
techniques~\cite{pichard,MacK} are then used to 
extract spectral information and linear conductance. Though simple enough, 
such model studies indeed bring out some subtleties of coherent spin 
dependent electronic transport, often showing the spin filtering effect 
over selected ranges of energy~\cite{maiti3}.

\textcolor{black}{Some recent studies on spin-based transistor~\cite{betthausen,saarikoski,wojcik} 
reveal the fact that spin transmission can be controlled by a 
suitable combination of a homogeneous magnetic field and a helical 
magnetic field in two-dimensional magnetic semiconductor waveguide 
structures. The relative strength of the homogeneous and helical
field components controls the backscattering process of the spins 
which changes conductance and the degree of spin polarization of 
transmitted electrons, and the device can be switched into `off' 
or 'on' state. In contrast to this, in the present study we try to 
explore the role of the local magnetic moments of the magnetic atomic 
sites in the chain to control the spin transmission for particles with 
spin $1/2$ as well as higher spin states in such model magnetic quantum 
device.}

While spin polarized transport of electrons has been the main 
concern so far, transport of particles with spins higher than that of 
an electron, even in one dimension (1D), has not received the same level 
of attention. This is, to our mind, an area which needs to be explored 
in order to unravel the possibility of designing novel storage devices 
which rely on the transportation or spin filtering of ultracold bosonic 
or fermionic quantum gases exhibiting higher spin states. Spin-$3/2$ 
particles, for example, can be realized with alkali atoms of $^6$Li, 
$^{132}$Cs, and alkaline earth atoms of $^9$Be, $^{135}$Ba, and 
$^{137}$Ba. These large-spin atomic fermions display diverse many-body 
phenomena, and can be made experimentally through controlled 
interactions in spin scattering channels~\cite{jiang}. Experimentally 
realized 1D strongly correlated liquids of ultracold fermions with a 
tunable number of spin components~\cite{pagano}, spin polarized hydrogen 
which remains gaseous down to zero temperature, and happens to be a good 
candidate for Bose condensation in a dilute atomic gas, or a 
gas of ultracold $^{52}$Cr atoms forming a dipolar gas of high spin 
atoms~\cite{fattori}, are some of the recently developed quantum systems 
which provide a versatile and robust platform for probing fundamental 
problems in condensed matter physics, as well as finding applications 
in quantum optics and quantum information processing.

Thus the availability of high spin state particles opens up an unexplored 
area of engineering spin filters for spins higher than $S=1/2$. It is 
already appreciated that, in contrast to the conventional 
spin-$1/2$ electronic case, large-spin ultracold atomic fermions, even 
in 1D, exhibit richer spin phenomena~\cite{jiang}. Thus, exploring the possibility 
of selecting out a state with a definite spin projection using a 
suitable quantum device might lead to innovative manipulation and 
control of spin transport. This precisely, is the motivation of the 
present communication.

We get exciting results. Using a simple 1D chain of magnetic atoms, 
mimicking a quantum gas in an artificial periodic potential, we show 
within a tight-binding framework that a suitable correlation between 
the spin $S$ of the incoming beam of particles, and the magnetic 
moment $\vec{h}$ offered by the substrate atoms can open up a gap in 
the energy spectrum. The opening of the gap turns out to be crucial 
in transporting a given spin state over a specified range of Fermi 
energy, while blocking the remaining spin states. The simple 1D chain 
of magnetic atoms of spin $S$ is shown to be equivalent to a 
$(2S+1)$-strand ladder network. This equivalence is exploited to 
work out the precise criterion of opening up of the spectral gap. 
The results seem to be robust against at least a minimal incorporation 
of disorder, as suggested by the results of spin polarized transport 
for a spin spiral, as reported in this communication.

In section~\ref{sec2} we describe the basic scheme in terms of 
the spin-$1/2$ and spin-$1$ particles. The difference equations are 
established which will finally be used to obtain the spectra in the 
respective cases. Section~\ref{sec3} describes how to engineer the 
spectral gaps using an appropriate substrate and provides the general 
criterion for opening up of the spectral gap for arbitrary spins. The 
discussion is substantiated by a detailed presentation of the density 
of states (DOS) choosing the spin-$1/2$ case as example. In section~\ref{sec4} 
we present the results of the transport calculations for the spin-$1/2$ 
and spin-$1$ cases. In section~\ref{sec5} we discuss the idea of having 
a spin filtering with a correlated disorder. In section~\ref{sec6} we show 
the robustness of the results by considering a spin-spiral where the 
substrate atoms have their magnetic moments turning sequentially, in a 
periodic fashion, mimicking disorder over a length shorter than the 
period, and in section~\ref{conclu} we draw our conclusions. The 
appendices contain the details of the transport formulation, and 
further details are provided in the supplemental material.
\section{Modelling a spin $S$ chain as a ladder of width $2S+1$}
\label{sec2}
\begin{figure}[tb]
\centering
\includegraphics[clip,width=\columnwidth, angle=0]{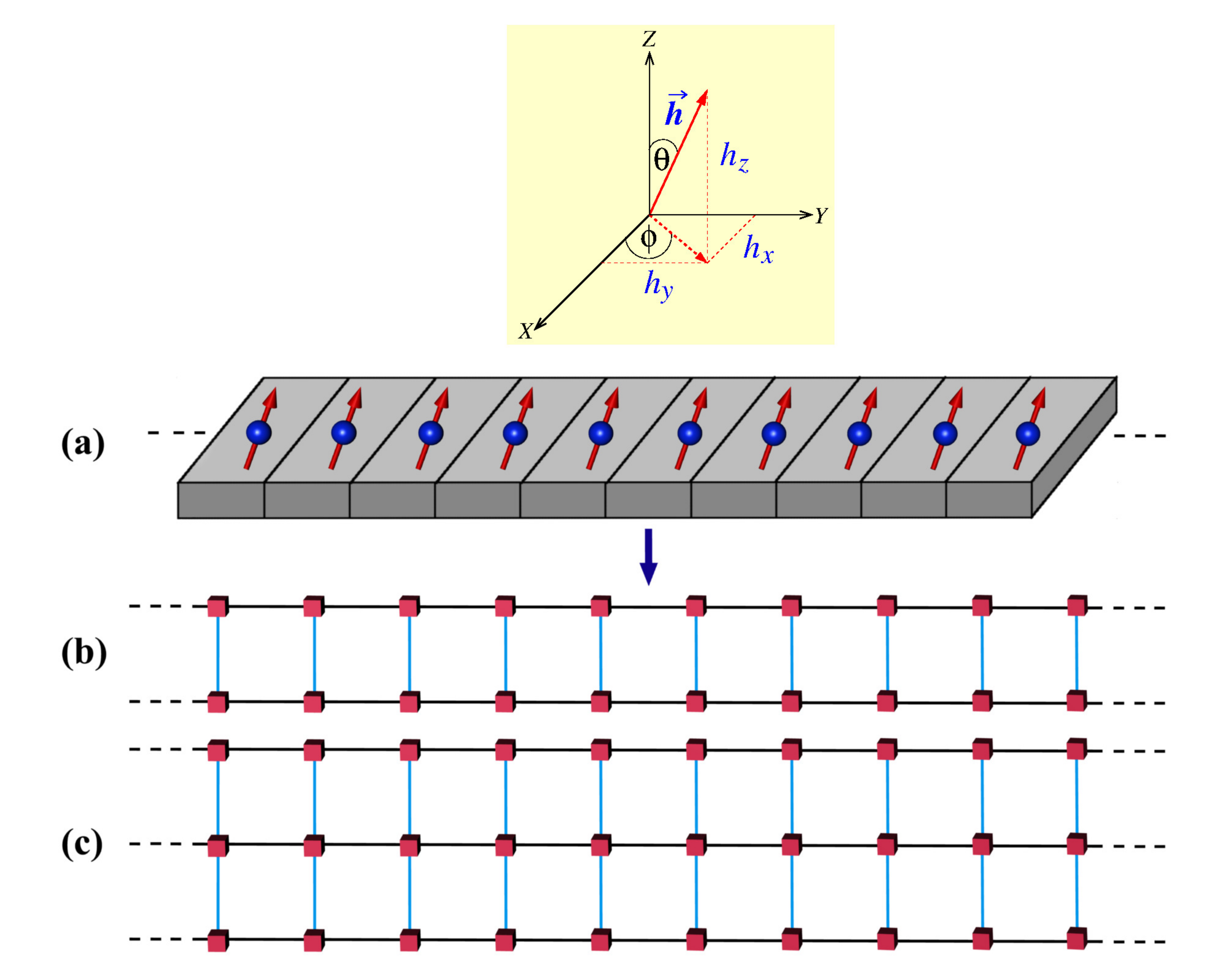}
\caption{(Color online) (a) Schematic diagram of a linear magnetic 
chain grafted on a substrate (gray blocks). Each magnetic atom 
(blue sphere) is subject to a substrate-induced magnetic moment $\vec{h}=(h_{x},h_{y},h_{z})$ 
(red arrow) making an angle $\theta$ with the $z$ axis and $\phi$ is 
the azimuthal angle. (b) Schematic representation of a two-strand 
ladder network with solid lines denoting the hopping elements and 
cubes representing the effective ``sites". (c) Schematic 
representation of a three-strand ladder network. The decomposition 
of $\vec{h}$ into its components is shown in the inset box above 
the magnetic chain. } 
\label{mapping}
\end{figure}

In Fig.~\ref{mapping} we propose a model
system consisting of a linear array of {\it magnetic} atoms grafted on 
a substrate. 
The atom 
at the $n$-th site has a magnetic moment $\vec{h}_n$ associated with it.  
The Hamiltonian in the tight-binding approximation can be written as
\begin{eqnarray}
\bm{H} & = & \sum_{n} \bm{c}_{n}^{\dagger} \left( \bm{\epsilon}_{n} - 
\vec{h}_{n} \cdot \bm{S}^{(S)}_{n} \right) \bm{c}_{n} \nonumber +\\ 
&  & \sum_{\langle n,m \rangle} \bm{c}_{n}^{\dagger} \bm{t}_{n,m} 
\bm{c}_{m} + \bm{c}_{m}^{\dagger} \bm{t}_{n,m} \bm{c}_{n},
\label{hamiltonian}
\end{eqnarray}
with $\langle n,m \rangle$ denoting nearest neighbors, \emph{i.e.}, $m=n\pm 1$. 
Each of the quantities 
$ \bm{c}_{n}^{\dagger}$, $\bm{c}_{n}$, $\bm{\epsilon}_{n}$, 
$\bm{t}_{n,m}$ and $ \bm{S}^{(S)}_{n}$ denotes a multi-component 
expression according to the spin content, \emph{i.e.}, for the $S=1/2$ case, 
$\bm{c}_{n}^{\dagger} = (c_{n,\uparrow}^{\dagger}, 
c_{n,\downarrow}^{\dagger})$ is the creation operator at the $n$th site, 
$\bm{\epsilon}_{n} = \mathrm{diag} \left( \epsilon_{n,\uparrow}, 
\epsilon_{n,\downarrow} \right)$ describes the diagonal on-site potential matrix, 
and $\bm{t}_{n,m} = \bm{t} = \mathrm{diag}\left( t, t \right)$ 
encodes the uniform nearest-neighbor hopping integral $t$ along $n$. 
The indices `$\uparrow$', `$\downarrow$' refer to the spin projections 
(spin `channels') for the case $S=1/2$. 

It is easily appreciated that the dimensions of the matrices increase 
proportionately as one extends the scheme to spin $1$, $3/2$ or higher values. 
Consequently, the creation and the annihilation operators will also have 
multiple components indexed by every single value of the spin projection 
$m_{S} = -S$, $-S+1$, $\ldots$, $S-1$, $S$, having a total of $2S+1$ values for 
a general spin-$S$ particle. 
The term $\vec{h}_{n} \cdot \bm{S}^{(S)}_{n} = h_{n,x} \bm{S}^{(S)}_{n,x} + 
h_{n,y} \bm{S}^{(S)}_{n,y} + h_{n,z} \bm{S}^{(S)}_{n,z}$ describes the interaction of 
the spin ($S$) of the injected particle with the localized on-site magnetic 
moment $\vec{h}_{n}$ at site $n$. This term is responsible for spin flipping 
at the magnetic sites. For $S=1/2$, $1$, $3/2$, $\ldots$, the $\bm{S}_{x}$, $\bm{S}_{y}$, $\bm{S}_{z}$ 
denote the generalized Pauli spin matrices $\bm{\sigma}_{x}$, $\bm{\sigma}_{y}$, $\bm{\sigma}_{z}$ 
expressed in units of $\hbar S$. Spin flip scattering is hence dependent on 
the orientation of the magnetic moments $\vec{h}_{n}$ in the magnetic chain 
with respect to the $z$ axis. 
Written explicitly for $S=1/2$ we have
\begin{eqnarray}
\vec{h}_{n} \cdot \bm{S}^{(1/2)}_{n} & = & h_{n,x}\bm{\sigma}_{x} + 
h_{n,y}\bm{\sigma}_{y} + h_{n,z}\bm{\sigma}_{z} \nonumber \\
& = & \left(\def\arraystretch{1.5}\begin{array}{cc}
h_{n} \cos\theta_{n} & h_{n} \sin\theta_{n} e^{-i\phi_{n}} \\
h_{n} \sin\theta_{n} e^{i\phi_{n}} & -h_{n} \cos\theta_{n} 
\end{array} \right),
\label{spinflip}
\end{eqnarray}
with $\theta_{n}$ and $\phi_{n}$ denoting polar and azimuthal angles, respectively.

The time-independent Schr\"{o}dinger equation for the 
pure spin-$1/2$ system is written as $H |\chi\rangle = E |\chi\rangle$, 
where $|\chi \rangle = \sum_{n} \big( \psi_{n,\uparrow} |n,\uparrow\rangle +
\psi_{n,\downarrow} |n,\downarrow\rangle \big)$ is a linear combination of 
spin-{\it up} ($\uparrow$) and spin-{\it down} ($\downarrow$) Wannier orbitals. 
Operating $H$ on $|\chi \rangle$ we get two equations relating the 
$\psi_{n,\uparrow}$, $\psi_{n,\downarrow}$ amplitudes on position $n$  
with the neighboring $n\pm 1$ sites,
\begin{subequations}
\label{spineq}
\begin{multline}
\left(E-\epsilon_{n,\uparrow} + h_{n}\cos \theta_{n} \right) \psi_{n,\uparrow} + 
h_{n} \sin\theta_{n} e^{-i\phi_{n}} \psi_{n,\downarrow} \\
= t \psi_{n+1,\uparrow} +
t \psi_{n-1,\uparrow},
\label{spineq1}
\end{multline}
\begin{multline}
\left(E-\epsilon_{n,\downarrow}-h_{n}\cos\theta_{n} \right) \psi_{n,\downarrow} +
h_{n} \sin\theta_{n} e^{i\phi_{n}} \psi_{n,\uparrow} \\
= t \psi_{n+1,\downarrow} + 
t \psi_{n-1,\downarrow}.
\label{spineq2}
\end{multline}
\end{subequations}
Eqs.~\eqref{spineq1} and \eqref{spineq2} can be expressed as 
matrix equation of the form,
\begin{equation}
( E \bm{1} - \tilde{\bm{\epsilon}}_{n} ) \bm{\psi}_{n} = 
\bm{t} \bm{\psi}_{n+1} + \bm{t} \bm{\psi}_{n-1},
\label{matdiffeqn}
\end{equation}
where
\begin{equation}
\tilde{\bm{\epsilon}}_{n}=
\left( \arraycolsep=5pt \def\arraystretch{1.5}\begin{array}{cccc}
\epsilon_{n,\uparrow}-h_{n} \cos\theta_{n} & -h_{n} \sin\theta_{n} e^{-i\phi_{n}} \\ 
-h_{n} \sin\theta_{n} e^{i\phi_{n}} & \epsilon_{n,\downarrow}+h_{n} \cos\theta_{n}
\end{array}
\right),
\label{effonsite-spinhalf}
\end{equation} 
and $\bm{\psi}_{n}= \left( \psi_{n,\uparrow}, \psi_{n,\downarrow} \right)$.
We draw the attention of the reader to the equivalence of Eq.~\eqref{spineq} 
to the difference equations for a {\it spinless} electron in a two-strand 
ladder network, as depicted in Fig.~\ref{mapping}(b) with an effective 
on-site potential $\epsilon_{n,\uparrow} - h_{n} \cos\theta_{n}$, and 
$\epsilon_{n,\downarrow} + h_{n} \cos\theta_{n}$ for the `upper' strand 
(identified with $\uparrow$ component) and the `lower' strand (identified with 
$\downarrow$ component) respectively. The amplitude of the hopping integral 
along each arm of the ladder is $t$, while $h_{n} \sin\theta_{n} \exp(i\phi_{n})$ 
plays the role of {\it inter-strand} hopping integral along the $n$-th 
strand~\cite{sil1}. 

Similarly, for spin $S=1$ particles, we have $2S+1=3$ coupled equations 
analogous to Eq.~\eqref{spineq}, namely,
\begin{widetext}
\begin{subequations}
\begin{align}
\left[E - (\epsilon_{n,1} - h_{n}\cos \theta_{n}) \right] \psi_{n,1} + 
\dfrac{1}{\sqrt{2}} h_{n}\sin \theta_{n}e^{-i\phi_{n}} \psi_{n,0} 
 =  t \psi_{n+1,1} + t \psi_{n-1,1}, & \\
\left[E - \epsilon_{n,0} \right] \psi_{n,0} +  
\dfrac{1}{\sqrt{2}} h_{n}\sin \theta_{n}e^{i\phi_{n}} \psi_{n,1} + 
\dfrac{1}{\sqrt{2}} h_{n}\sin \theta_{n}e^{-i\phi_{n}} \psi_{n,-1} 
 =  t \psi_{n+1,0} + t \psi_{n-1,0}, & \\
\left[E - (\epsilon_{n,-1} + h_{n}\cos\theta_n) \right] \psi_{n,-1} + 
\dfrac{1}{\sqrt{2}} h_{n}\sin \theta_{n}e^{i\phi_{n}} \psi_{n,0} 
 =  t \psi_{n+1,-1} + t \psi_{n-1,-1}. &
\end{align}
\label{eqspin1}
\end{subequations}
\end{widetext}
for the three spin projections, viz., $+1$, $0$ and $-1$.
\begin{figure*}[tb]
\centering
(a)\includegraphics[clip,width=0.6\columnwidth, angle=0]{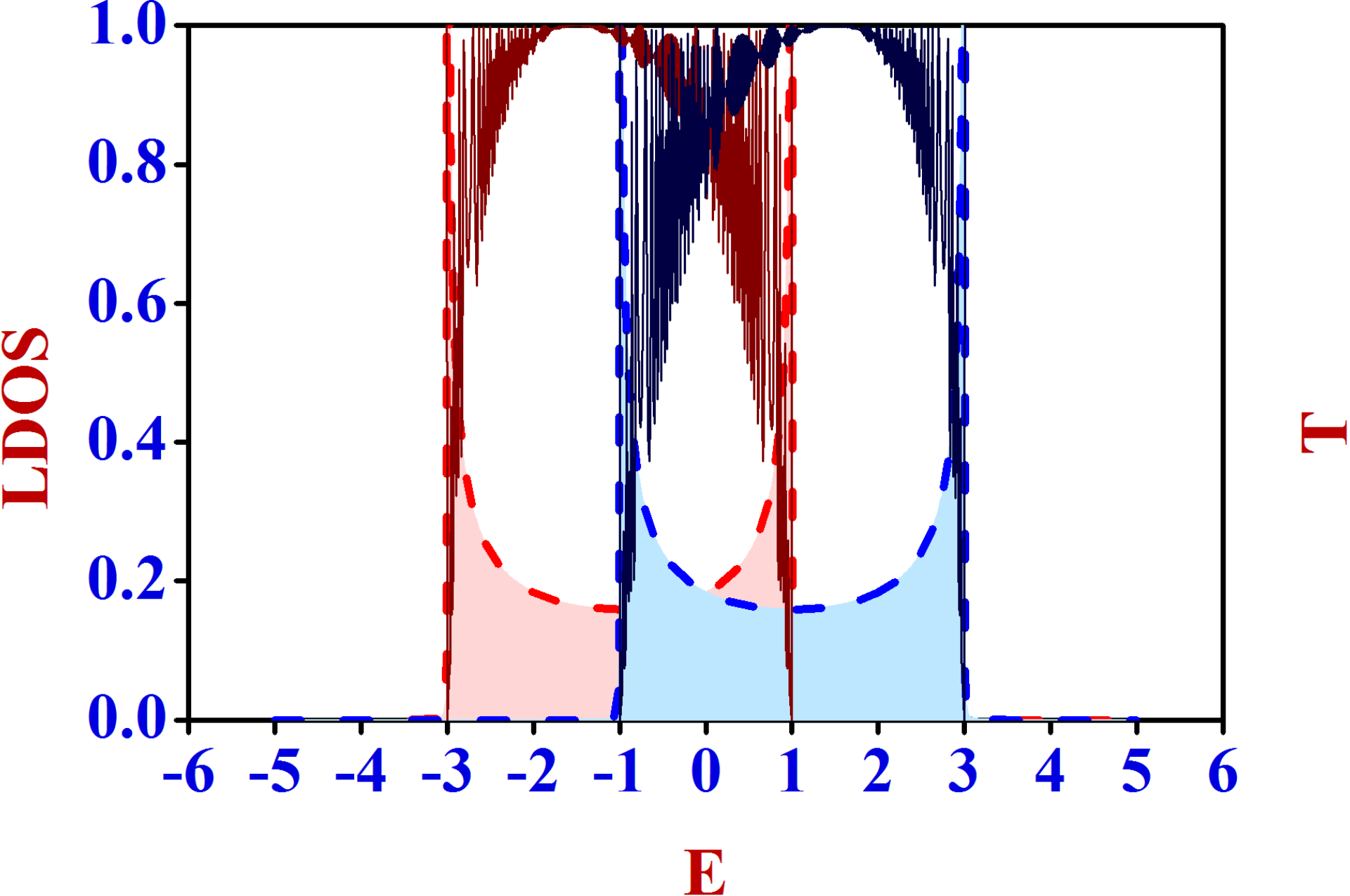}
(b)\includegraphics[clip,width=0.6\columnwidth, angle=0]{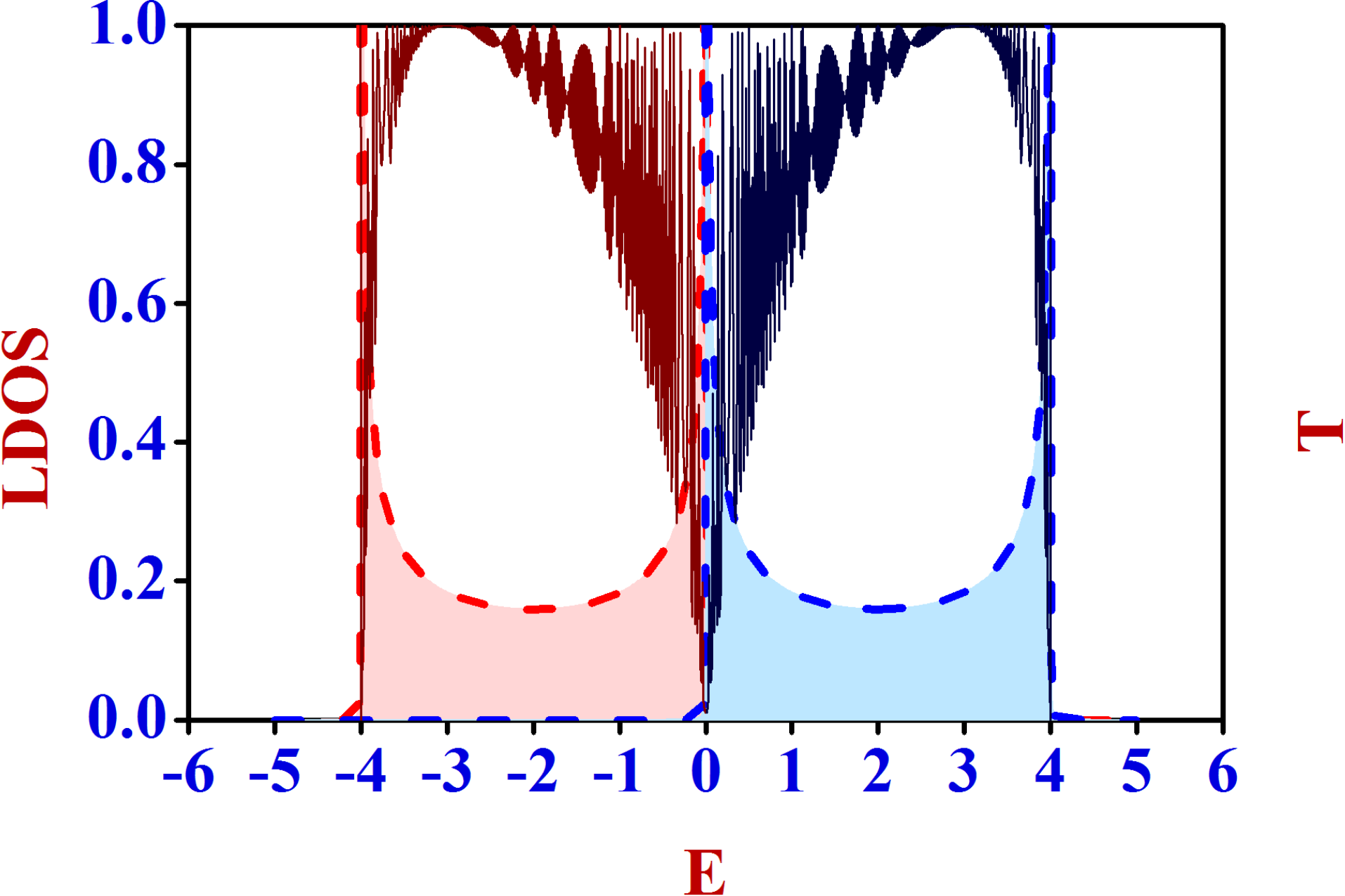}
(c)\includegraphics[clip,width=0.6\columnwidth, angle=0]{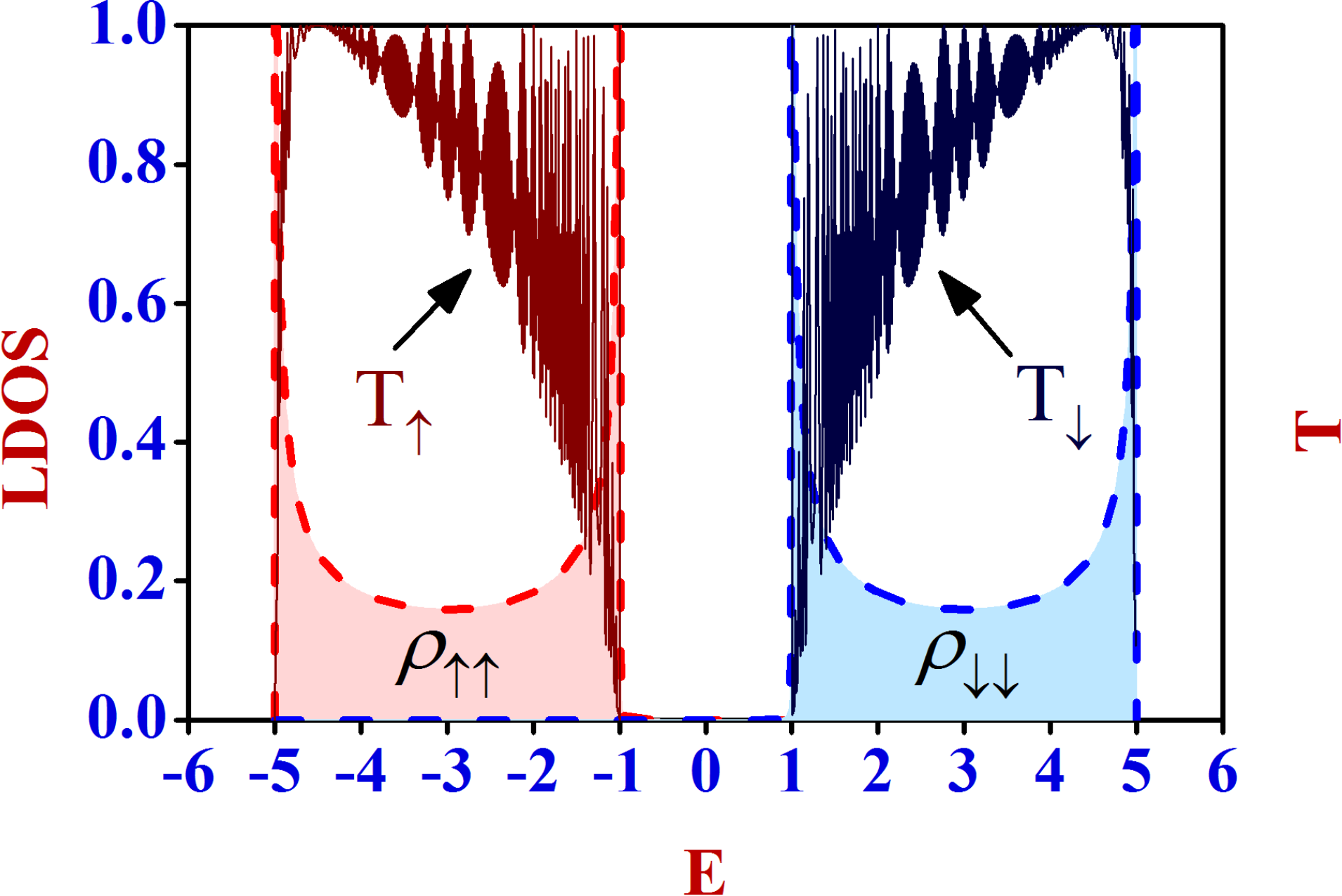}
\caption{(Color online) Variation of local density of states (LDOS) 
and transmission probabilities 
with energy $E$ as a function of $h$ for a fixed value of $\theta=0$ for 
spin-$1/2$ particles. The light red shaded plot with red envelope is the 
LDOS for the spin-up ($\uparrow$) states and the light blue shaded plot 
with blue envelope is the LDOS for the spin-down ($\downarrow$) states. 
The dark red curve represents the transmission characteristics for the spin-up ($\uparrow$) particles and the dark blue curve is exhibiting the transport 
for the spin-down ($\downarrow$) particles. (a) is for $h=1$, (b) is for 
$h=2$, and (c) corresponds to $h=3$. The lead parameters for the 
non-magnetic leads are $\epsilon_{L}=\epsilon_{R}=0$ and 
$t_{LD}=t_{RD}=t_{L}=t_{R}=3$.} 
\label{ldos-trans-half-theta0}
\end{figure*}
%
The $3 \times 3$ matrix for the `effective' on-site potential at the $n$-th 
position now reads,
\begin{equation}
\tilde{\bm{\epsilon}}_{n} = 
\left( \arraycolsep=5pt \def\arraystretch{1.5}\begin{array}{ccccc}
\epsilon_{n,1}-\lambda_{n} & -\dfrac{1}{\sqrt{2}}\xi_{n} e^{-i\phi_{n}} & 0\\ 
-\dfrac{1}{\sqrt{2}}\xi_{n} e^{i\phi_{n}} & \epsilon_{n,0} 
& -\dfrac{1}{\sqrt{2}}\xi_{n} e^{-i\phi_{n}}\\
0 & -\dfrac{1}{\sqrt{2}}\xi_{n} e^{i\phi_{n}} & \epsilon_{n,-1}+\lambda_{n}
\end{array}
\right),
\label{effonsite-spin1}
\end{equation}
where $\lambda_{n}=h_{n} \cos\theta_{n}$ and $\xi_{n}=h_{n} \sin\theta_{n}$.
Clearly, this can be extended to treat the case of general $S$, leading to 
an effective ladder model with $2S+1$ arms as shown in Fig.~\ref{mapping}.
The above equations, viz., Eq.~\eqref{spineq} and Eq.~\eqref{eqspin1} will now 
be exploited to engineer the spectral gaps and simulate spin filters, as 
explained in the subsequent sections.
\section{Engineering the spectral gaps}
\label{sec3}
\subsubsection{Computing the density of states}
Its instructive to remind ourselves as to when one can have a gap in 
the spectrum of such a ladder network in terms of the simplest possible 
case, where one can set $\epsilon_{n,\uparrow}=\epsilon_{n,\downarrow}=
\epsilon_{n}=\epsilon$, a constant at all sites 
$n$ of both the arms of the ladder, and $h_{n} = h$. We additionally set 
$\theta_{n} = \theta$ and $\phi_{n}=0$. It is easy to understand that, 
in the extreme limit of $t\rightarrow 0$, the spectrum of the two-strand 
ladder yields sharply localized (pinned) eigenstates at $E = \epsilon \pm h$. 
The \textcolor{black}{DOS} will exhibit two $\delta$-function spikes at 
these energy eigenvalues. As the hopping along the arms of the ladder, viz., 
$t$ is switched `on', the $\delta$-function-like spikes in the DOS 
spectrum broaden into two subbands, which will finally merge into a single 
band when $h\sim t$. Therefore, for a given value of the polar angle $\theta$, 
and a predefined value of $t$ (which sets the scale of energy), the 
inter-strand hopping $h$ can be tuned to open or close a gap in the 
energy spectrum.

Mapping back onto the original 1D magnetic chain the above 
argument clearly shows that one can create gaps in the spectrum or close 
them, by a judicious engineering of the substrate, that is, the required 
species of the magnetic atoms providing an appropriate value of the magnetic 
moment $h$. This simple argument allows us to gain analytical control 
over the spectrum and eventually turns out to be crucial in designing a 
spin filter. In Fig.~\ref{ldos-trans-half-theta0} we show the DOS of a uniform 
magnetic chain with $\theta=0$, and $\phi=0$. The DOS for the `up' and 
the `down' spin electrons in the magnetic chain with $\epsilon=0$ and $t=1$ 
have been calculated by evaluating the matrix elements of the Green's 
function $\bm{G} = (E \bm{1} - \bm{H})^{-1}$ in the Wannier basis 
$|j,\uparrow (\downarrow) \rangle$. The local DOS (same as the average DOS 
in this case) for the `up' and `down' spin electrons are given by, 
\begin{equation}
\rho_{\uparrow\uparrow (\downarrow\downarrow)} = \lim_{\eta \rightarrow 0} 
\langle j,\uparrow(\downarrow)|\bm{G}(E+i\eta)| 
j,\uparrow(\downarrow)\rangle .
\label{green}
\end{equation} 
Here, $\rho_{\uparrow\uparrow}$ and $\rho_{\downarrow\downarrow}$ have been 
evaluated using a real space decimation renormalization method elaborated 
elsewhere~\cite{arunava,LeaRS99a}.

\subsubsection{Substrate-induced opening and closing of spectral gaps}
The choice of the strength of the magnetic moment that will make the 
spectrum gapless is not quite arbitrary. One can, at least for a special 
relative orientation of the moments at the nearest neighboring sites, 
work out a prescription for this. To appreciate the scheme, let us 
observe that, even for a site dependent potential $\epsilon_n$ and 
the magnetic moment $h_{n}$, the commutator $[\tilde{\bm{\epsilon}}_{n},
\tilde{\bm{\epsilon}}_{n+1}] = 0$ if we choose 
\textcolor{black}{$\theta_{n+1} - \theta_{n} = m\pi$, 
where $m = 0,\,\pm 1,\,\pm 2,\,\pm 3,\,....$} and 
$\phi_{n}=\phi_{n+1}=0$ or a constant value, 
irrespective of the values of $\epsilon_{n}$ or $h_{n}$. That is, the system may 
represent either a ferromagnetic alignment of the moments, or an antiferromagnetic one. 
Needless to say, the specific case of constant $\epsilon$ and constant $\theta$ 
falls in this category. In such cases, it is possible to decouple the matrix 
equation~\eqref{matdiffeqn} into a set of two independent linear equations by 
making a change of basis, going from $\tilde{\bm{\psi}}_{n}$ to 
$\bm{\mathcal{F}}_{n}$, where, $\bm{\mathcal{F}}_{n} = 
\bm{S}^{-1}\tilde{\bm{\psi}}_{n}$. $\bm{S}$ executes a similarity 
transformation on Eq.~\eqref{matdiffeqn}. The commutation ensures that 
every $\tilde{\bm{\epsilon}}_{n}$ matrix can be diagonalized 
simultaneously by the same matrix $\bm{S}$. The concept has previously 
been used to study the electronic spectrum of disordered and quasiperiodic 
ladder networks~\cite{sil1,sil2,pal}, and 2D lattices with correlated 
disorder~\cite{alberto}. 
\begin{figure*}[tb]
\centering
(a)\includegraphics[clip,width=0.6\columnwidth, angle=0]{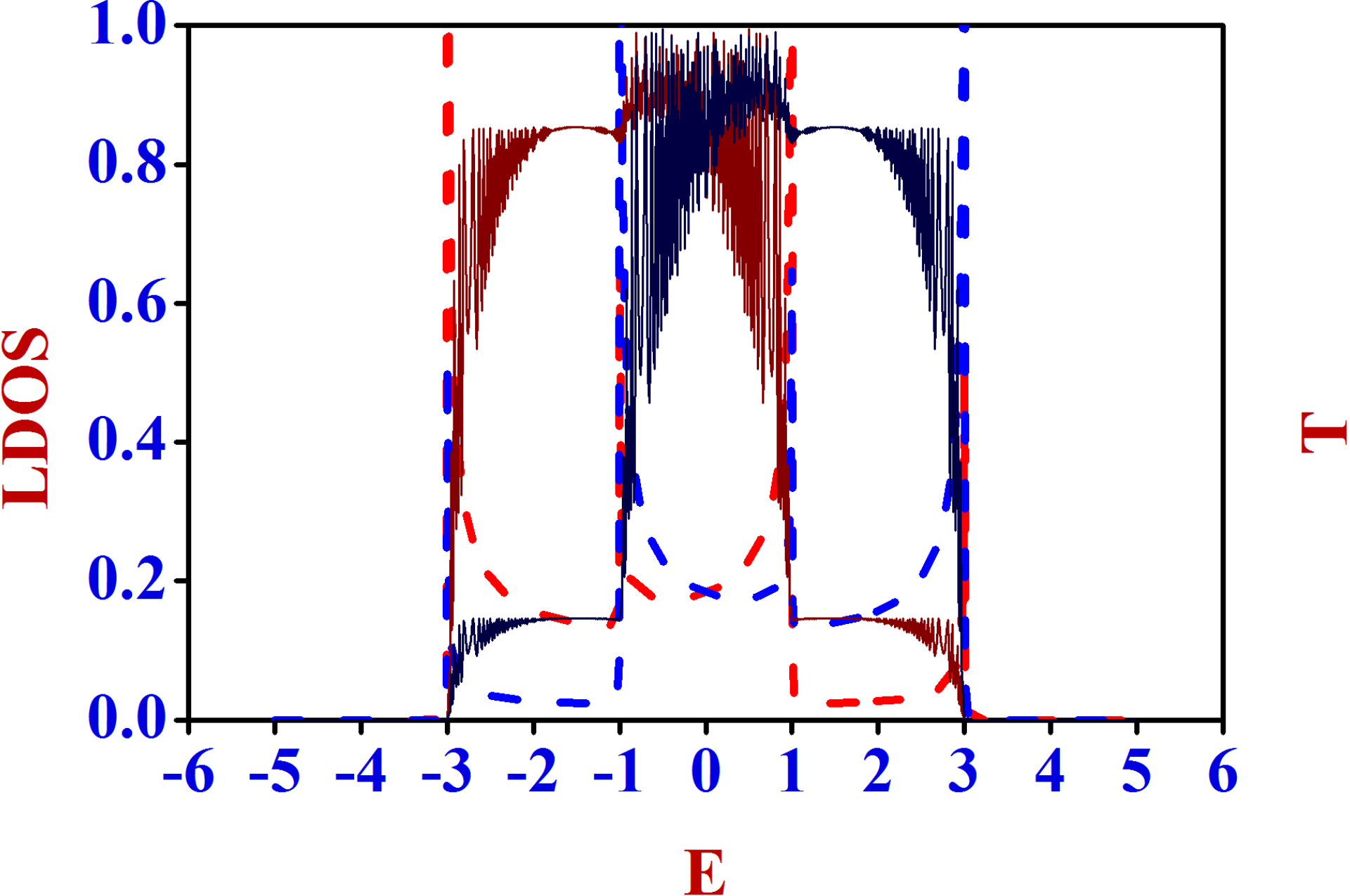}
(b)\includegraphics[clip,width=0.6\columnwidth, angle=0]{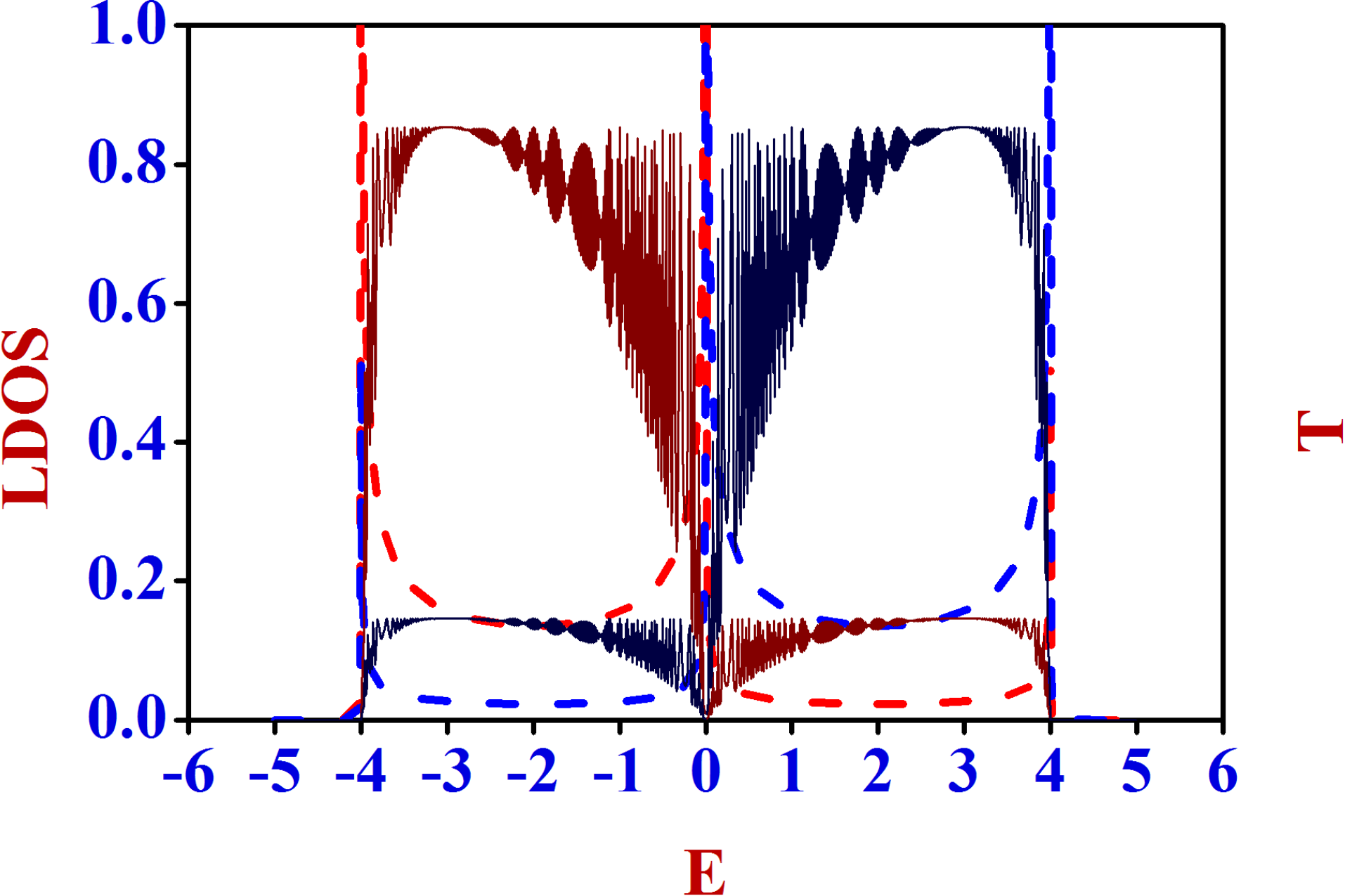}
(c)\includegraphics[clip,width=0.6\columnwidth, angle=0]{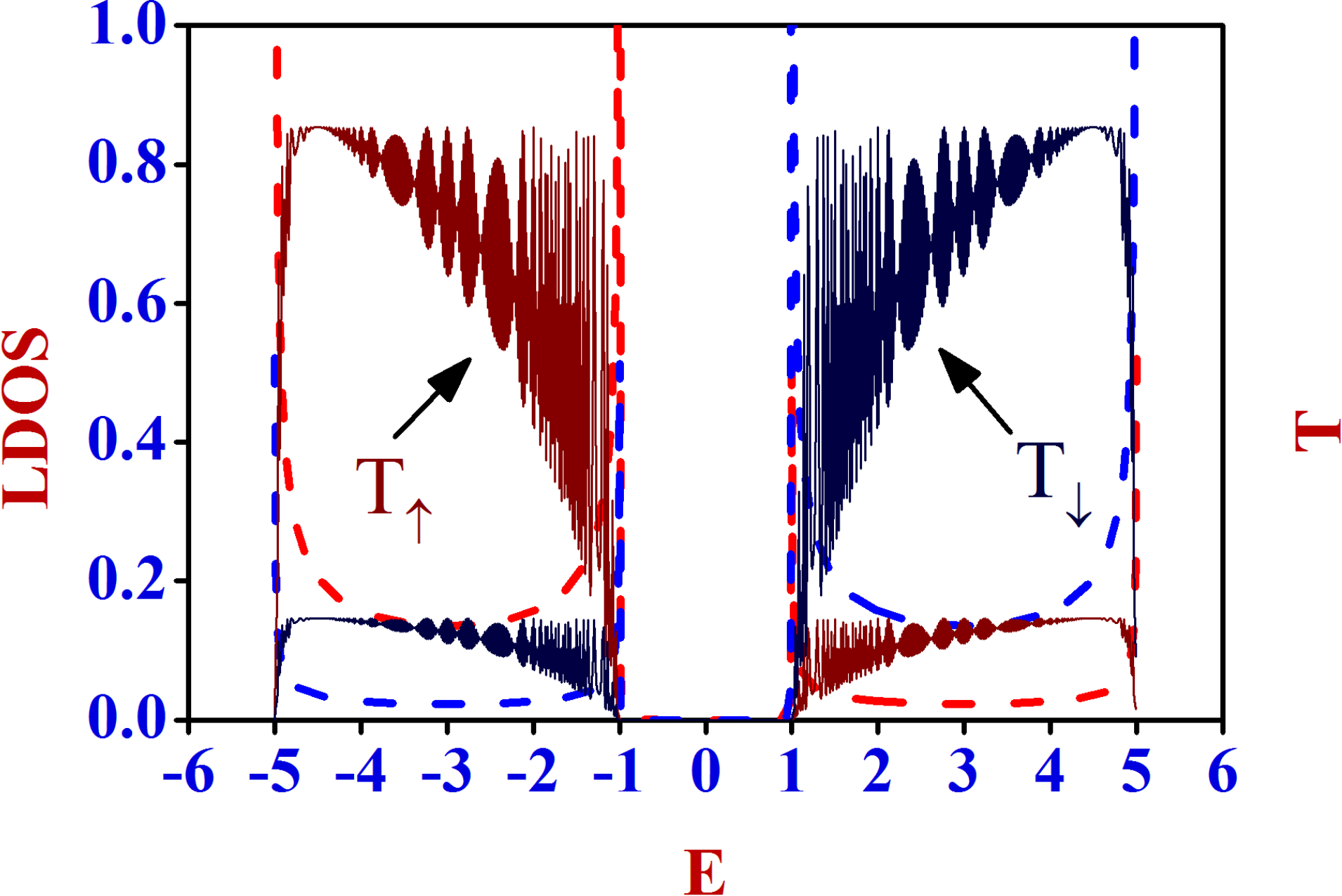}
\caption{(Color online) Variation of LDOS and transmission probabilities with 
energy $E$ as a function of $h$ for a fixed value of $\theta=\pi/4$ for spin-$1/2$ particles. The curves with the red dashed line represent the LDOS for 
the spin-up ($\uparrow$) states and the curves with the blue dashed line 
represent the LDOS for the spin-down ($\downarrow$) states. The dark red 
curve represents the transmission probability for the spin-up ($\uparrow$) 
particles and the dark blue curve is exhibiting the transmission probability 
for the spin-down ($\downarrow$) particles. (a) is for $h=1$, 
(b) is for $h=2$, and (c) corresponds to $h=3$.} 
\label{ldos-trans-half-thetaPiby4}
\end{figure*}
The decoupled equations read,   
\begin{subequations}
\begin{eqnarray}
\left(E-\epsilon_{n} + h_{n}\right) \mathcal{F}_{n,\uparrow} = 
t \mathcal{F}_{n+1,\uparrow} + t \mathcal{F}_{n-1,\uparrow},
\label{spliteq1}\\
\left(E-\epsilon_{n} - h_{n}\right) \mathcal{F}_{n,\downarrow} = 
t \mathcal{F}_{n+1,\downarrow} + t \mathcal{F}_{n-1,\downarrow},
\label{spliteq2}
\end{eqnarray}
\label{spliteq}
\end{subequations}
and represent the equations for two {\it pseudoparticles} with mixed spin states,
\begin{subequations}
\begin{eqnarray}
\mathcal{F}_{n,\uparrow} & = & \frac{\sin \theta_{n}}{2} \psi_{n,\uparrow} + 
\frac{1-\cos \theta_{n}}{2} \psi_{n,\downarrow}, \\
\mathcal{F}_{n,\downarrow} & = & -\frac{\sin \theta_{n}}{2} \psi_{n,\uparrow} + 
\frac{1+\cos \theta_{n}}{2} \psi_{n,\downarrow}.
\label{pseudo}
\end{eqnarray} 
\end{subequations}
The subscripts `$\uparrow$' or `$\downarrow$' in $\mathcal{F}_{n}$ can be 
taken to be the indices for the two decoupled arms of the equivalent 
two-strand ladder.
 
Let us again get back to the perfectly ordered case with $\epsilon_{n} = 
\epsilon$, and $h_{n} = h$. The first thing to appreciate is that each 
individual equation, viz., Eq.~\eqref{spliteq1} and 
Eq.~\eqref{spliteq2} now represents a perfectly periodic array of atomic 
like sites with an {\it effective} on-site potential $\epsilon \pm h$, 
and a constant nearest neighbor hopping integral $t$. Consequently they 
offer {\it absolutely continuous} energy bands, ranging from 
$\epsilon - h -2t$ to $\epsilon - h +2t$ corresponding to  
Eq.~\eqref{spliteq1}, and $\epsilon + h -2t$ to $\epsilon + h +2t$ for the 
second decoupled equation Eq.~\eqref{spliteq2}. 
The energy spectrum for the individual infinite 
chains can be obtained conventionally by working out the DOS for each of 
them. The DOS of the actual linear magnetic chain is then obtained through a 
convolution of these two individual DOS's. It is simple to compute that the 
gap between the bands is given by, 
\begin{equation}
\Delta = 2 h - 4 t.
\end{equation}
This immediately leads to a {\it critical} value of the strength of the 
magnetic moment $h_{c} = 2t$ for which the gap will {\it just} close. 
The result is independent of any arbitrary constant value of the polar 
angle $\theta$, as long as one ensures that the difference between nearest 
neighboring values of the angle, viz., 
\textcolor{black}{$\theta_{n+1} - \theta_{n} = m\pi$ 
($m = 0,\,\pm 1,\,\pm 2,\,\pm 3,\,....$)}.

The variation of the DOS against energy as a function of $h$ is a generic 
feature of the magnetic array for any constant value of the polar angle 
$\theta$. This is evident from  Figs.~\ref{ldos-trans-half-theta0} and 
\ref{ldos-trans-half-thetaPiby4}, where we have plotted the 
local density of states (LDOS) at a site in the infinite chain. 
For a periodic chain the LDOS is same as the average DOS. 
It is clear from the figures, how a 
gradual increase in the value of the magnetic moment $h$, a gap opens in 
the spectrum, going through a sequence of variations shown in each panel, 
where the values of $\rho_{\uparrow\uparrow}$ and $\rho_{\downarrow\downarrow}$ 
complement each other. \textcolor{black}{To be noted that, in all the plots the 
energy $E$ in the abscissa is taken in units of $t$}.

\subsubsection{Spectral gaps for larger $S$}
\begin{figure*}[ht]
\centering
(a)\includegraphics[clip,width=0.6\columnwidth, angle=0]{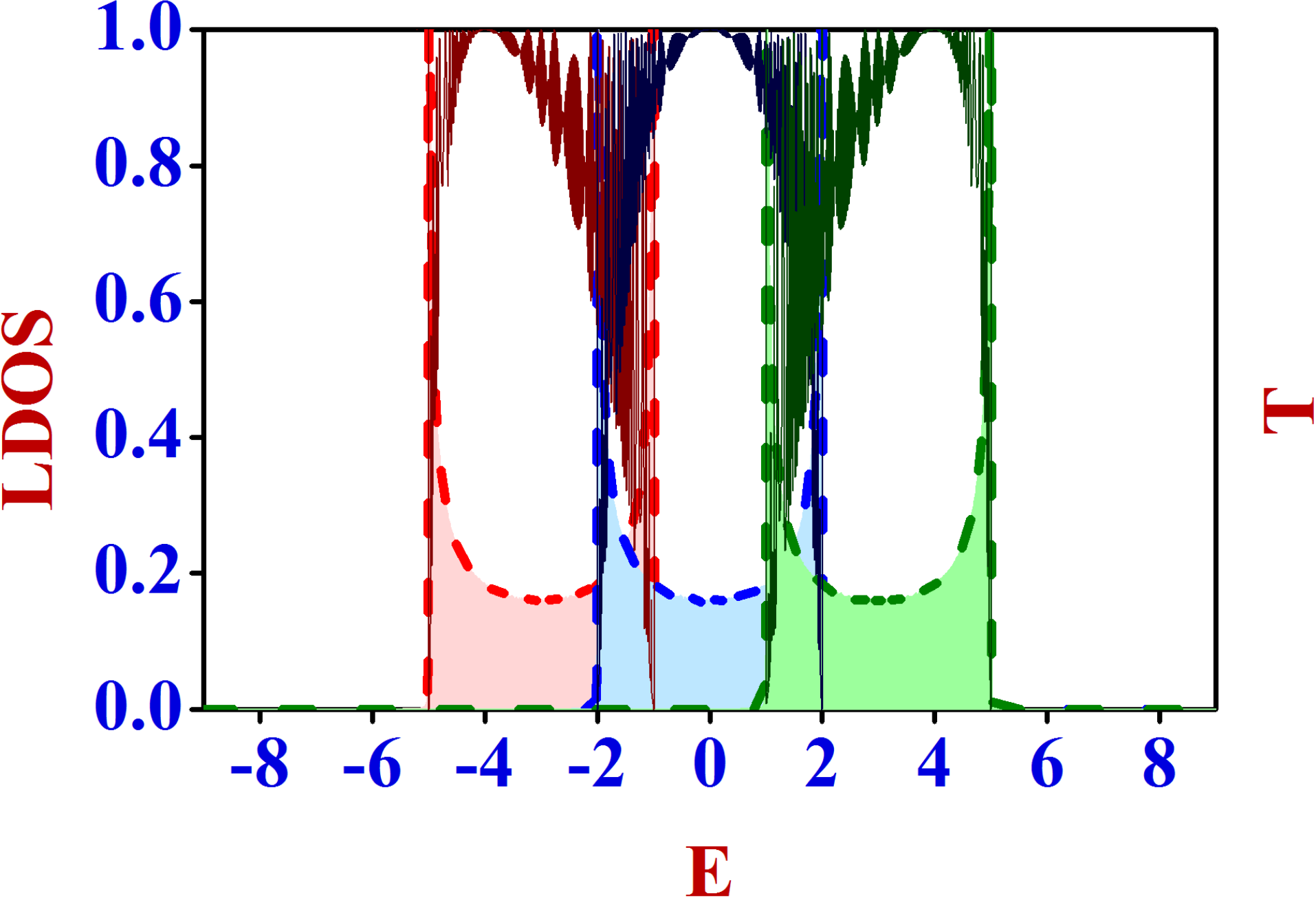}
(b)\includegraphics[clip,width=0.6\columnwidth, angle=0]{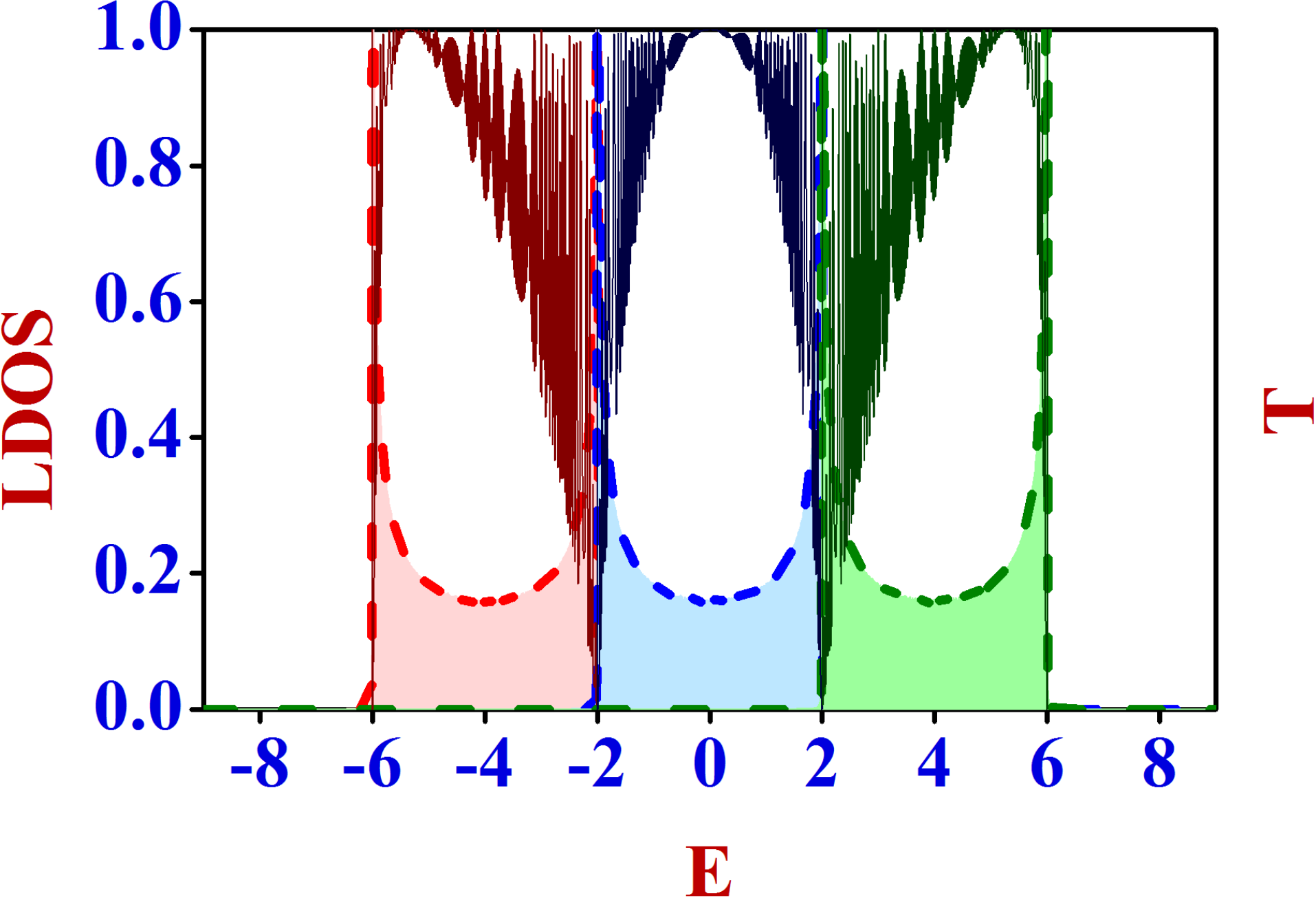}
(c)\includegraphics[clip,width=0.6\columnwidth, angle=0]{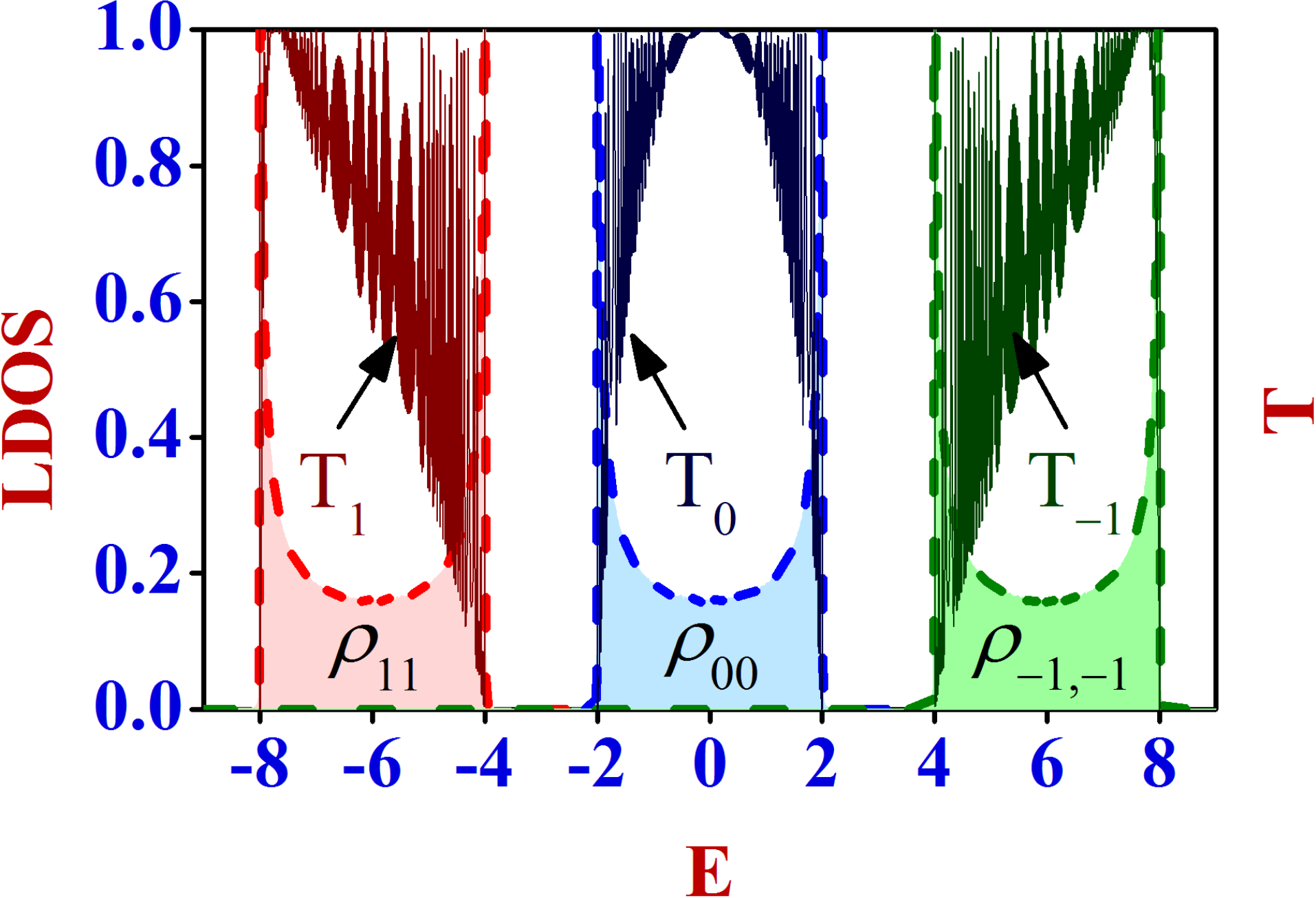}
\caption{(Color online) Variation of LDOS and transmission probabilities 
with energy $E$ as a function of $h$ for a fixed value of $\theta=0$ for 
spin-$1$ particles. The light red shaded plot with the dashed red envelope 
is the LDOS for the spin-$1$ states, the light blue shaded plot with the 
blue dashed envelope is the LDOS for the spin-$0$ states, and the light 
green shaded plot with the green dashed envelope is the LDOS for the spin-$(-1)$ 
states. The dark red curve represents the transmission characteristics 
for the spin-$1$ particles, the dark blue curve is for the spin-$0$ particles, 
and the dark green curve stands for the spin-$(-1)$ particles. (a) is for 
$h=3$, (b) is for $h=4$, and (c) corresponds to $h=6$. The lead parameters 
for the non-magnetic leads are $\epsilon_{L}=\epsilon_{R}=0$ and 
$t_{LD}=t_{RD}=t_{L}=t_{R}=4$.} 
\label{ldos-trans-1-theta0}
\end{figure*}

A look at the set of Eq.~\eqref{eqspin1} immediately reveals the equivalence 
of the magnetic chain in the present case with that of a three-strand ladder 
as depicted in Fig.~\ref{mapping}(c). The effective on-site potential at 
the $n$-th vertex at every strand is given by $\epsilon_{n} - h_{n} \cos\theta_n$, 
$\epsilon_{n}$ and $\epsilon_{n} + h_{n} \cos\theta_n$ respectively, while the 
role of the inter-strand coupling (hopping integral) between the adjacent strands 
is played by $h_{n} \sin\theta_n/\sqrt{2}$ (with $\phi_{n}$ is set equal zero). 
As before, one can argue that an 
appropriate tuning of $h_{n}$ (for a given value of $\theta_{n} = \theta =$ 
constant) should open up gaps in the spectrum, in the same way as 
it did in the spin-$1/2$ case. This is precisely what we see in 
Fig.~\ref{ldos-trans-1-theta0}. 
We have chosen a chain with a constant value of the on-site potential 
$\epsilon_{n} = \epsilon$, $h_{n}$ has been fixed to any desired constant 
value, and $\theta_{n} = \phi_{n} = 0$. The left, middle and the bottom panels 
exhibit the overlap of bands for $h = 3$, the marginal case where the bands 
just touch each other for $h = 4$, and a clear opening of the gaps when 
$h = 6$, respectively. The location of the gaps can be estimated quite easily 
if one observes that with $\theta = 0$, the strands in the three-arm ladder 
effectively get decoupled, so that one is left with a set of three independent 
equations representing three individual ordered chains with on site potentials 
$\epsilon - h$, $\epsilon$, and $\epsilon + h$ respectively. The corresponding 
ranges of eigenvalues are, $[\epsilon -h -2t, \epsilon -h +2t,]$, $[\epsilon -2t,
\epsilon +2t]$, and $[\epsilon +h -2t,\epsilon +h +2t]$. The gap between these 
ranges can now be estimated in a straightforward way, and the critical value 
of $h$, for which gaps will open for any spin $S$ can thus be worked out to be, 
\begin{equation}
\Delta^{(S)} = \frac{h}{S} - 4 t.
\label{gapeq}
\end{equation} 
This equation holds for any spin, viz., $S = 1/2$, $1$, $3/2$, $2$, $5/2$, $\ldots$. 
So in principle, we can engineer the bands corresponding to the different 
components of any higher spin. The results for spin $3/2$ and spin $2$ are 
shown in the supplemental material~\cite{supplement}. \textcolor{black}{The basic 
mechanism of designing spin filters by controlling the value and the orientation 
of the magnetic moments of magnetic atoms in the chain remains the same as we move 
on to the higher spin states. For the higher spin states the number of spin channels 
increases, and we have a whole lot of options as to which spin component we want 
to make transmitting through the system for a certain energy regime. We need to 
tune the magnitude of $h$ accordingly to have a spin filtering effect as we climb 
up to the higher spin states.}   
\section{Two-terminal transport and spin filtering}
\label{sec4}
We now discuss the results of two-terminal transport across the 
magnetic chain. The detailed formulation of the method is provided in 
the Appendix. For simplicity, we plot the transmission coefficient in 
each case in the same figure as the corresponding LDOS.
We start with the simple case of spin $1/2$ and $\theta=0$, and choose 
values of $h$ such that there are (a) overlapping, (b) touching and 
(c) well-separated subbands. From Fig.~\ref{ldos-trans-half-theta0}(a) 
we see that the DOS of the `up' and the `down' spin states for $h=1$ 
overlap over one third of the range of allowed eigenvalues. The 
corresponding transmission spectrum naturally offers a mixed character. 
There is a partial filtering effect with only `up' spin electrons 
emerging out of the system in the energy interval $-3 \le E \le -1$, 
while it is the opposite in its positive counterpart. 
With $h = h_{c} = 2$, the subbands for the `up' and `down' spin states 
just touch each other. It is obvious from Fig.~\ref{ldos-trans-half-theta0}(b) 
that only `up' spin electrons get transported in the lower half of 
the band, i.e., in the range $-4 \le E \le 0$, while the `down' spin 
electrons transmit in the range $0 \le E \le 4$. With $h = 3$, the gap 
is explicit, and the spin filtering effect is clear. In this simple case, 
there is no `spin flip' effect, and the `up' (`down') electrons get 
transmitted precisely in the energy intervals in which the respective 
bands are populated. 
For $\theta=\pi/4$ the situation is more complicated as shown in 
Fig.~\ref{ldos-trans-half-thetaPiby4}. Here we see that for the same 
values of $h$ as in Fig.~\ref{ldos-trans-half-theta0}, we now always 
get transport with mixed spin-up and spin-down components, even when 
there is a well-pronounced gap at shown in 
Fig.~\ref{ldos-trans-half-thetaPiby4}(c). Adjusting the value of 
$\theta$ therefore allows to control the relative admixture of the 
transported spin states while the choice of $h$ determines the energy 
range in which the different transport channels will be open.

The behaviour shown for spin-$1/2$ persists for spin-$1$ (and higher 
spin states \cite{supplement}) as shown Fig.~\ref{ldos-trans-1-theta0} 
(a), (b) and (c) for $\theta=0$. The gap opens at a larger value of $h$, 
as compared to the spin-$1/2$ case and can be easily estimated from 
Eq.~\eqref{gapeq}. The DOS corresponding to the three spin components, 
namely, $1$, $0$ and $-1$, exhibit overlap in the panel (a) for $h=3$. 
The critical value of $h=4$ makes the subbands touch each other at the 
appropriate energy values, while a clean gap opens up as $h=6$. 
Consequently, the two terminal transport exhibits a partial filtering 
in selected energy regimes in panel (a) while there is perfect filtering 
in (b) and (c). We note in (a) that the partial filtering in transmission 
exists only between the spin channels $(1,0)$ and $(0,-1)$. 
The results can be understood intuitively if one recalls that $S=1$ is 
identical to the case of a three strand ladder. Since we are working with 
nearest neighbor hopping in both the longitudinal and the transverse 
directions and rigid boundary conditions in the $y$-direction, the upper 
arm of the ladder (equivalent to the spin-$1$ channel) is totally decoupled 
from the lower arm (the spin-$(-1)$ channel). The central arm, namely, 
the spin-$0$ channel is coupled to these two outer arms. The partial 
filtering is thus caused by contributions coming from the {\it 
interacting pair} of `arms', or equivalently, the spin channels. 
Clearly, this argument extends also to the higher spin states.
\section{Spin filtering with correlated disorder}
\label{sec5}
\begin{figure}[tb]
\centering
(a)\includegraphics[clip,width=0.6\columnwidth, angle=0]{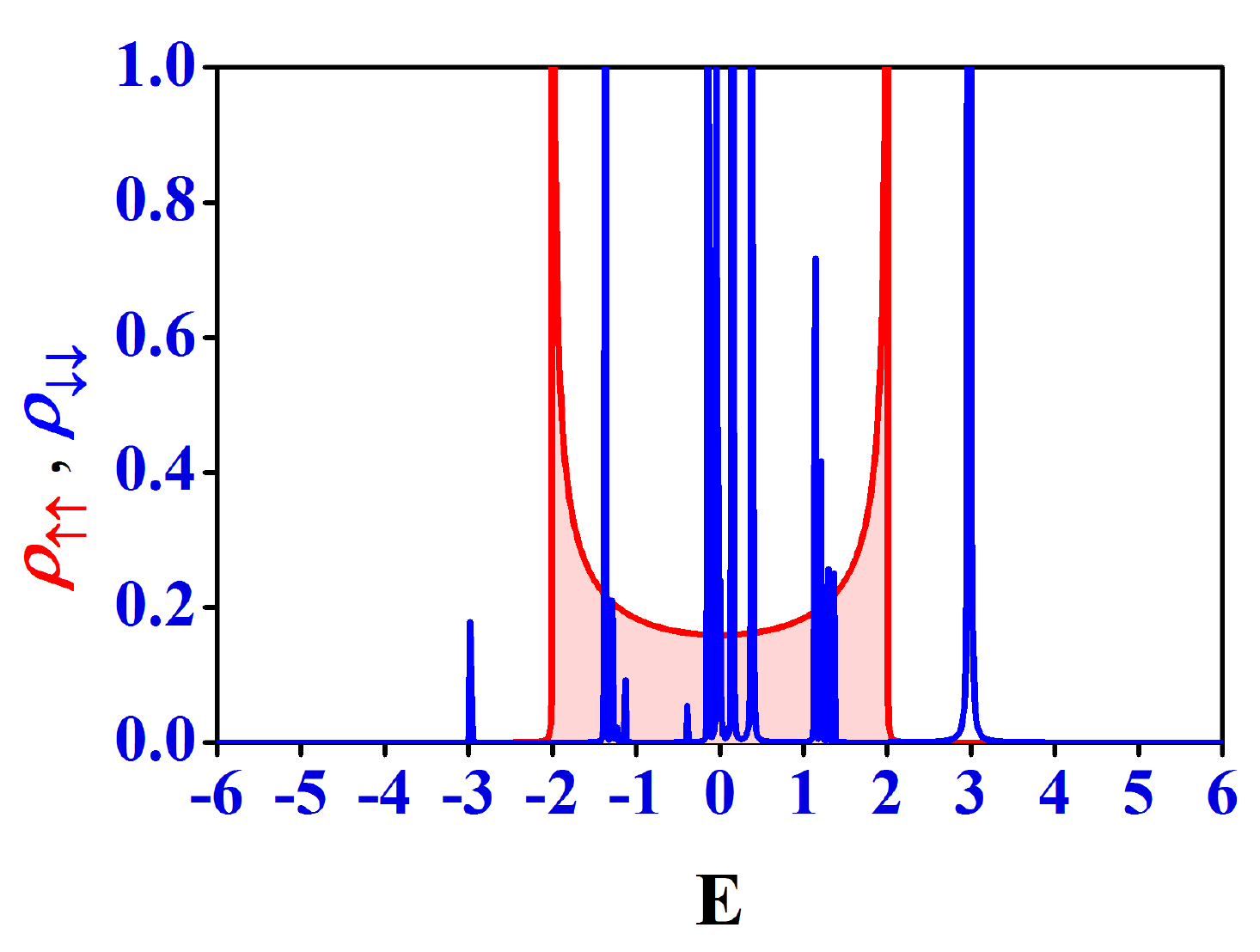}\\
(b)\includegraphics[clip,width=0.6\columnwidth, angle=0]{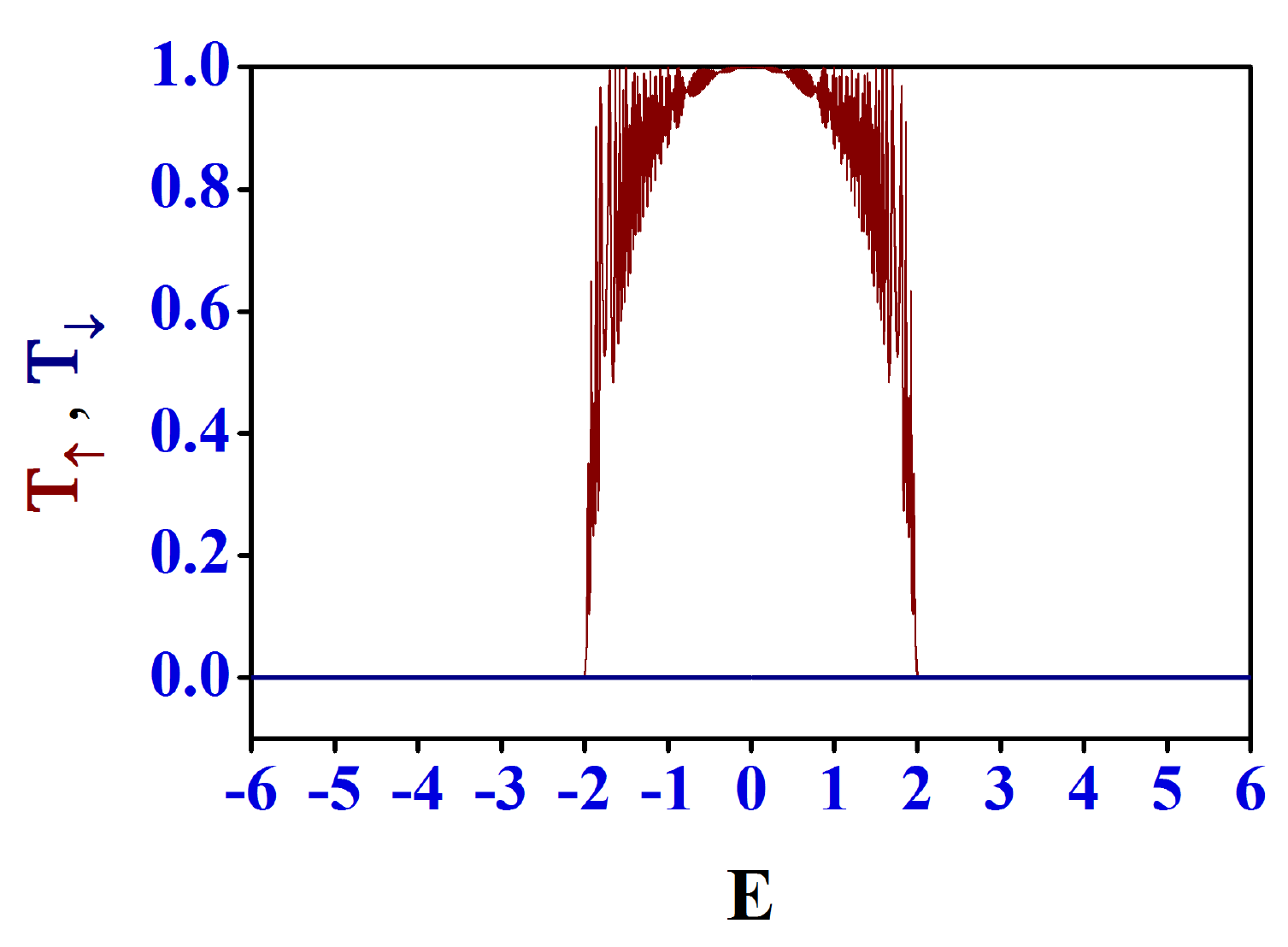}
\caption{(Color online) (a) Densities of states of the decoupled set of 
equations~\eqref{spineq-theta0} when $\epsilon_{n} = h_{n} = \mu \cos(\pi Q n a)$ with $\mu=1$ 
and $Q = (\sqrt{5}+1)/2$. The red and the blue lines correspond to the 
Eq.~\eqref{spineq1-theta0} and Eq.~\eqref{spineq2-theta0} respectively. 
(b) The zero transmission for the `down' spins reflect the `critical' character 
of the wavefunctions obtained from Eq.~\eqref{spineq2-theta0}, and the high 
transmission of the `up' spins reveal the extended nature of the 
wavefunctions obtained from Eq.~\eqref{spineq1-theta0}, and represents 
a perfect spin filter. We have set $t=1$, and energy is measured in 
units of $t$. The lattice constant $a=1$.} 
\label{ldos-trans-half-A-A}
\end{figure}
A perfect spin filter can be designed even without setting 
constant values for the on-site potentials and the substrate 
magnetic moments, and without bothering about engineering gaps 
in the energy spectrum. This can be achieved by introducing 
correlations between the numerical values of $\epsilon_{n,\uparrow} = 
\epsilon_{n,\downarrow} = \epsilon_{n}$ and the magnitude of the 
substrate moments $h_n$. We demonstrate a special situation in 
Fig.~\ref{ldos-trans-half-A-A}. For simplicity, but without 
sacrificing the central spirit, we assign an Aubry-Andre 
variation~\cite{aubry} in the on-site potential, viz., 
$\epsilon_{n} = \mu \cos(\pi Q n a)$ with $Q=(\sqrt{5}+1)/2$. 
This distribution of the on-site potential leads to extended, 
critical or localized eigenstates for $\mu < 2t$, $\mu = 2t$ 
and $\mu > 2t$ respectively~\cite{aubry}. The distribution of the magnetic moments 
$h_{n}$ are chosen to be equal to $\epsilon_{n}$. In addition, we 
set $\theta_{n}=0$. With the choice of $\theta_{n}=0$, and $\epsilon_{n,\uparrow} 
= \epsilon_{n,\downarrow} = \epsilon_{n}$, the set of equations~\eqref{spineq1} 
and~\eqref{spineq2} map into the following set of equations,
\begin{subequations}
\begin{eqnarray}
\left(E-\epsilon_{n} + h_{n}\right) \psi_{n,\uparrow} = 
t \psi_{n+1,\uparrow} + t \psi_{n-1,\uparrow},
\label{spineq1-theta0}\\
\left(E-\epsilon_{n} - h_{n}\right) \psi_{n,\downarrow} = 
t \psi_{n+1,\downarrow} + t \psi_{n-1,\downarrow}.
\label{spineq2-theta0}
\end{eqnarray}
\label{spineq-theta0}
\end{subequations} 
With $\epsilon_{n} = h_{n}$, the `effective' on-site potential in 
Eq.~\eqref{spineq2-theta0} is equal to $2\mu \cos(\pi Q n a)$, 
and a selection of $\mu=1$ makes the eigenstates corresponding to  
Eq.~\eqref{spineq2-theta0} {\it critical}. Eq.~\eqref{spineq1-theta0}, 
with $\epsilon_{n}=h_{n}$, now  represents a perfectly ordered chain 
with its spectrum ranging from $E = -2t$ to $E = 2t$. The densities 
of states and corresponding transport are shown in 
Fig.~\ref{ldos-trans-half-A-A}. In panel (a), the densities of states 
for the two decoupled channels are shown. The \textcolor{black}{DOS} for 
the original system is obtained by convolution of these two, but 
will definitely encompass the same energy regimes. The `up' spins 
now have an absolutely continuous \textcolor{black}{DOS} shown by the 
red shaded curve, ranging between $[-2t,2t]$. The critical eigenstates 
for the `down' spins are shown by the blue lines. As critical and 
extended states cannot coexist at the same energy, the central 
part of the spectrum will remain extended in the final, convolved 
\textcolor{black}{DOS}. The outer peaks however, will be there.

The corresponding transport characteristics are represented in panel
(b). The total `down' spin transport is naturally blocked, and 
we now have a clear case of spin filtering even with correlated, but 
a {\it deterministic disorder}, as is evident from the high transmission 
of `up' spin states between $[-2t,2t]$. It can be easily understood that, 
with a different choice of correlation between $\epsilon_{n}$ and $h_{n}$ 
(say, $\epsilon_{n} = - h_{n}$), 
we can make the `down' spin channel to be perfectly conducting, and the 
`up' spins to be completely blocked. As we go to the higher spin cases, the 
same trick can be used to make one of the spin channels to be perfectly 
conducting and the rest to be completely blocked. Obviously we need to have 
different correlations between $\epsilon_{n}$ and $h_{n}$ for different cases 
as we move along the higher spin ladder.    

It is obvious that we need not stick to the case of deterministic disorder only. 
If we choose both $\epsilon_{n}$ and $h_{n}$ in a random yet in correlated way, 
such that $\epsilon_{n} - h_{n} = \Lambda$ remains a  constant, i.e., 
$n$-independent, then Eq.~\eqref{spliteq1} yields an absolutely 
continuous spectrum in the range $\Lambda-2t \leq E \leq \Lambda + 2t$. 
This will be true even when the constant value of the polar angle $\theta \ne 0$.
All eigenstates in this energy range have to be of extended Bloch 
functions. On the other hand, even with this choice, Eq.~\eqref{spliteq2} 
represents a randomly disordered chain of scatterers for which the {\it 
pseudoparticle states} with mixed spin status will be Anderson localized. The 
system will then open up a transmitting channel for such {\it mixed spin states} 
only in the window $\Lambda-2t \leq E \leq \Lambda + 2t$, while it will remain 
opaque to all incoming electrons, irrespective of their spin states, 
in the energy regime beyond these limits \footnote{The same argument 
can of course be made when choosing  $\epsilon_{n} + h_{n} = \Lambda$. 
Then Eq.~\eqref{spliteq2} yields an absolutely continuous spectrum and 
Eq.~\eqref{spliteq1} represents a randomly disordered chain. Of course, 
the pertinent issue in this case is not filtering out a particular spin state.
Instead, one can address a serious issue of localization to delocalization 
crossover for a pseudoparticle of a mixed spin state.}. 

The above argument holds, and the scenario may even become richer, as 
probes with higher spin states are incident on the magnetic substrate. 
For a total spin $S$, with the same restriction on the polar angle 
$\theta_{n}$, and the azimuthal angle $\phi_{n}$ being set equal to zero, 
the matrix equation Eq.~\eqref{matdiffeqn} decouples into a set of 
$2S+1$ independent equations, each representing a pseudoparticle with 
a mixed spin state (now much more complicated). One can then introduce 
a correlation between $\epsilon_{n}$ and $h_{n}$, keeping them individually 
random, so as to make any one of these independent equations (say, the 
central one) represent a perfectly ordered linear chain with its band 
ranging between two energy values dictated by the effective on-site 
potential in that equation. The remaining $2S$ equations represent 
disordered linear chains with all pseudoparticle states exponentially 
localized. The spectra arising out of these $2S$ linear chains have 
their own `band centers' and can come arbitrarily close to the central 
continuum. The full spectrum is expected to be a dense packing of point-like distributions on either side of the central continuum.
Considering the overall charge transport, such cases may 
even give rise to the possibility of a metal-insulator transition. 
The situation is to be contrasted with the case of a real multiple 
stranded ladder network~\cite{alberto} remembering that, here we have 
just a single magnetic chain. \phantom{It is the spin state of the incoming 
particle that will decide whether such a simple quantum device will 
show up any reentrant behavior in charge transport or not.}
\section{A spin spiral: simulating `local' disorder}
\label{sec6}
In this section we extend the concepts developed in the earlier sections 
to a patterned magnetic chain mimicking a spin spiral~\cite{enkovaara} 
in one dimension. 
\begin{figure}[ht]
\centering
\includegraphics[clip,width=\columnwidth, angle=0]{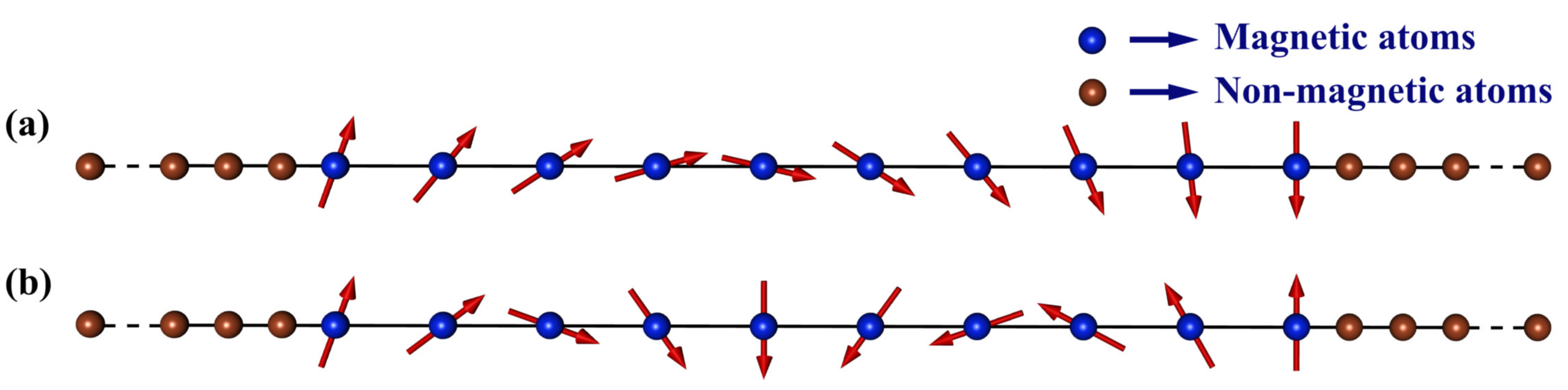}
(c) \quad $\theta_{n}=n \pi/L$ \hfill $\theta_{n}=2 n \pi/L$ \quad \mbox{ }\\
\includegraphics[clip,width=\columnwidth, angle=0]{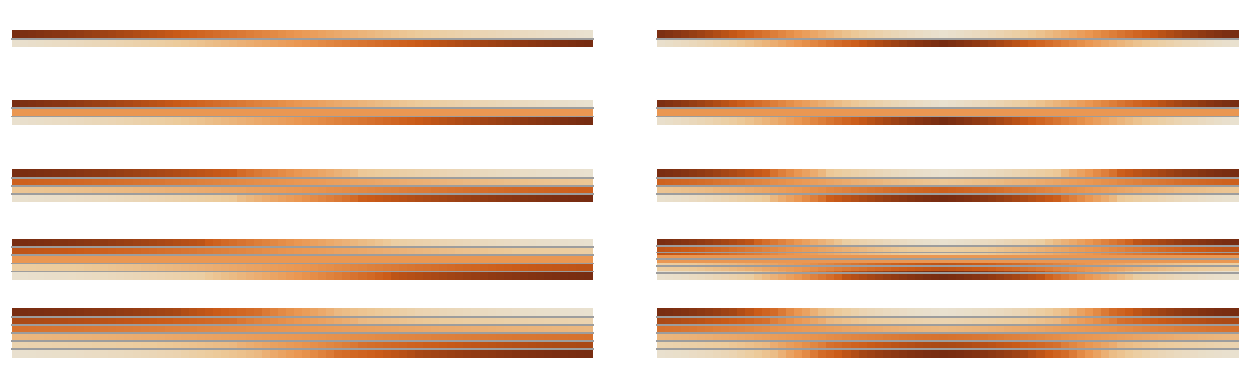}
\caption{(Color online) Schematic diagram of linear array of magnetic 
atoms (blue spheres) with magnetic moment vectors $\vec{h}$ (red arrows) 
forming spiral configurations for (a) $\theta_{n}=n \pi/L$ and (b) 
$2 n \pi/ L$. The system is connected between two non-magnetic 
semi-infinite leads (brown spheres). The horizontal lines are guides 
to the eye only.
(c) Color representation of the on-site potential $h\cos\theta_{n}$ of 
spiral configurations at $L=72$ for (left) $\theta_{n}=n \pi/L$ and 
(right) $2 n \pi/ L$ corresponding to $S=1/2$, $1$, $\ldots$, $5/2$ 
from top to bottom. Dark colors correspond to large values.} 
\label{spiralchain}
\end{figure}
The $n$-th atomic site now has its magnetic moment tilted 
with respect to the global magnetization axis (the $z$-axis) by an angle 
$\theta_{n}$. As we neglect any spin-orbit interaction in this work, the 
spin and the position spaces are decoupled, and the relative orientation 
of the neighboring spin becomes important in respect of the transport and 
other physical properties. Here we stick to a periodic variation of the 
spiral, though the period can be quite arbitrary. The configuration is 
schematically depicted in Fig.~\ref{spiralchain} and can be identified 
as a {\it frozen} magnon. If we keep our attention confined to the 
`chemical unit cell' only, the spiral configuration breaks the translational 
order locally (though preserving it globally of course) and simulates 
the effect of a kind of (deterministic) disorder, in particular, when 
the length of the chain is shorter than the period of the 
spin spiral. A study of the transmission 
characteristics for a spiral patch with length restricted to less than a 
period or its integral multiple, may show up some character expected for 
a real disordered magnetic chain, and test the robustness of the results 
obtained earlier.

We follow the same RSRG decimation scheme used earlier to study the LDOS at a bulk 
site for $\theta_{n} = n \pi/L$ and $\theta_{n} = n\pi/(L/2)$. Even when the 
strength of the magnetic moment $h$ is set to be same at every magnetic site, 
the variation in $\theta_{n}$ naturally leads to variations in the values of 
$h \cos\theta_{n}$ and $h \sin\theta_{n}$. This implies that we have a magnetic 
chain of atoms with a (deterministic) fluctuation both in the effective on 
site potential, viz., $\epsilon \pm h\cos\theta_{n}$ and the `coupling' 
$h \sin\theta_{n}$ between the `up' and the `down' spin channels. Mapping 
into the effective multi-strand ladder network, we have a case of a ladder 
where the `rungs' are associated with varying hopping integrals, simulating 
a kind of deformation, and where, at the same time the vertices are occupied 
by atomic sites with sequentially changing on-site potentials, see 
Fig.~\ref{spiralchain}(c).
\begin{figure*}[tb]
\centering
(a)\includegraphics[clip,width=0.6\columnwidth, angle=0]{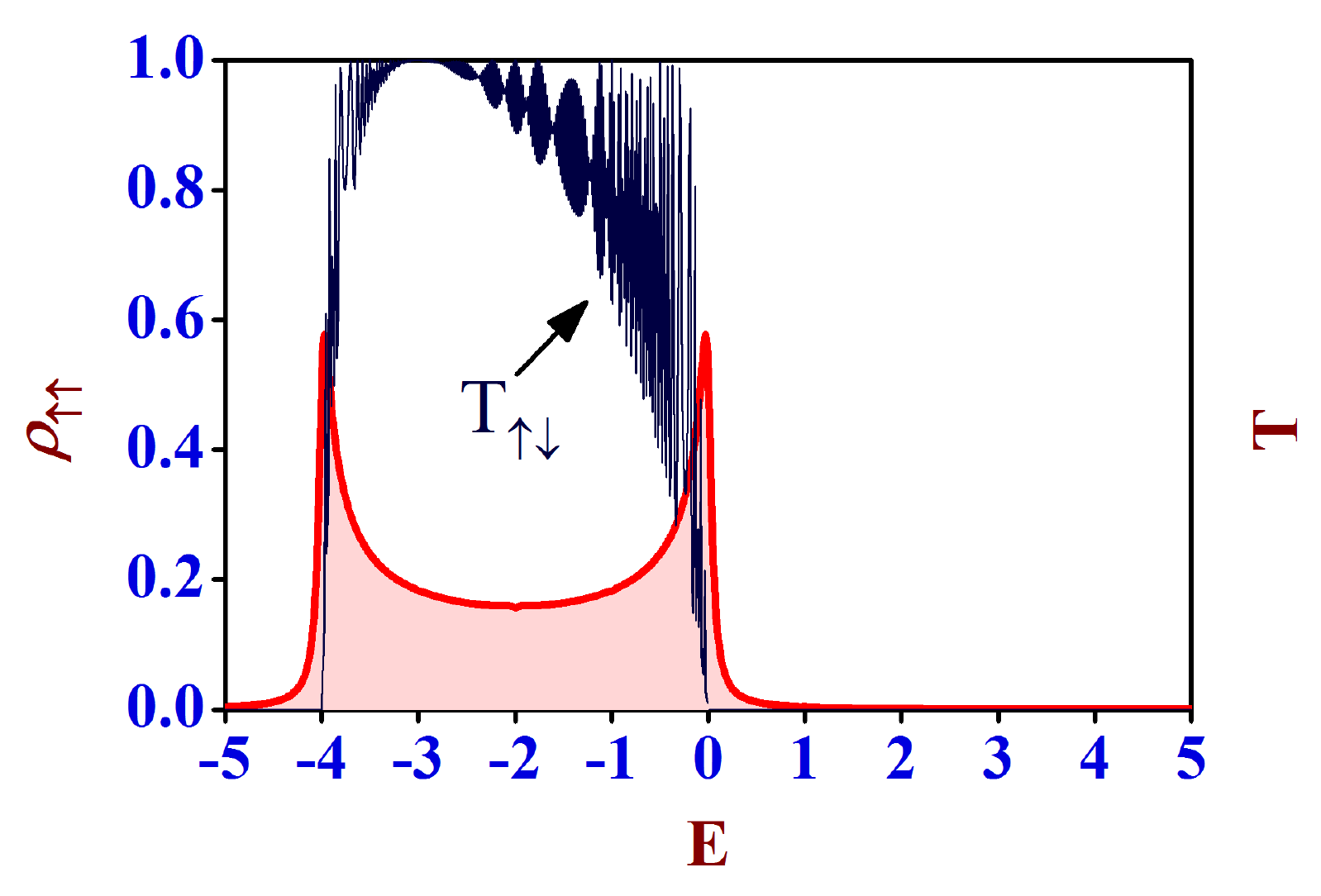}
(b)\includegraphics[clip,width=0.6\columnwidth, angle=0]{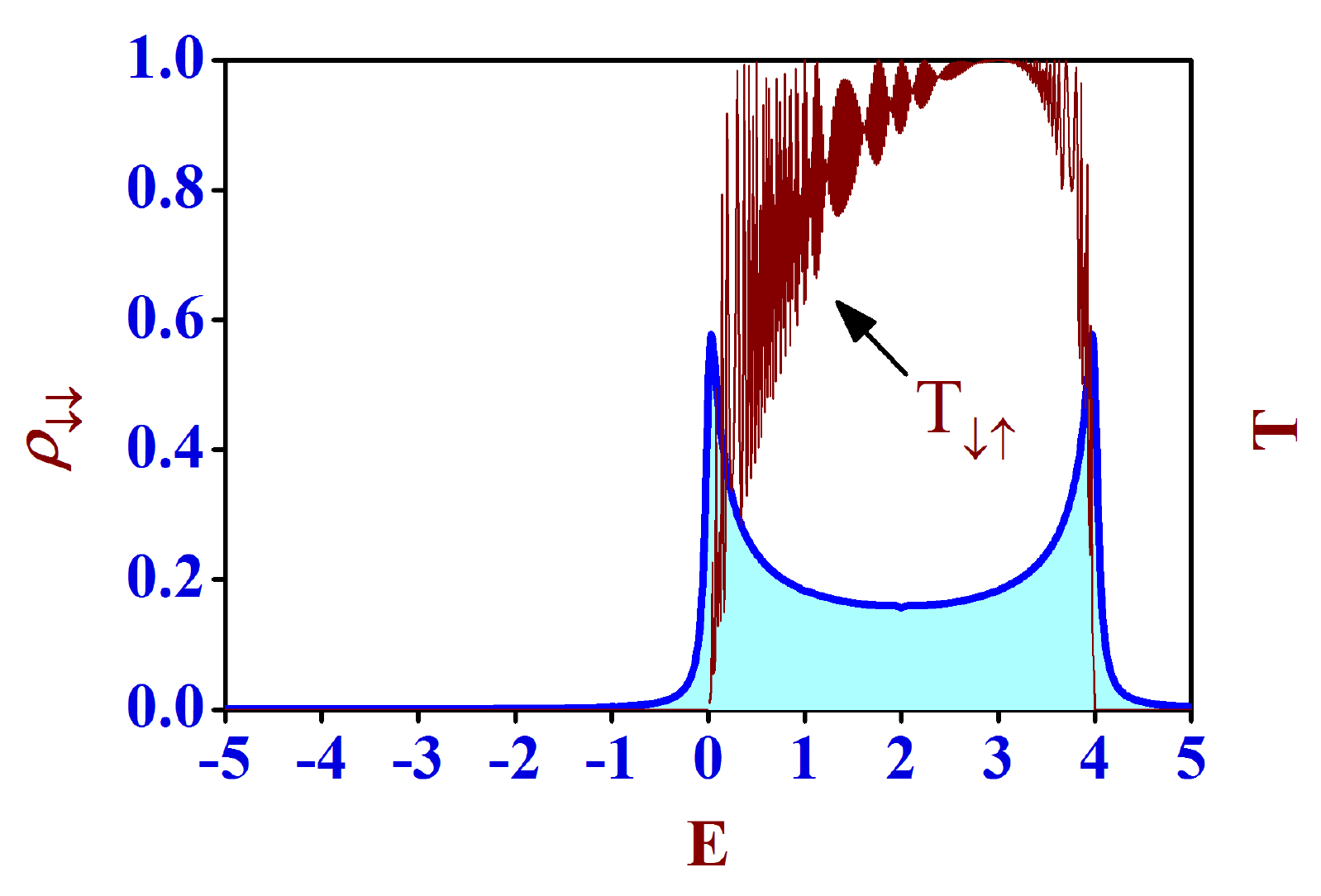}\\
(c)\includegraphics[clip,width=0.6\columnwidth, angle=0]{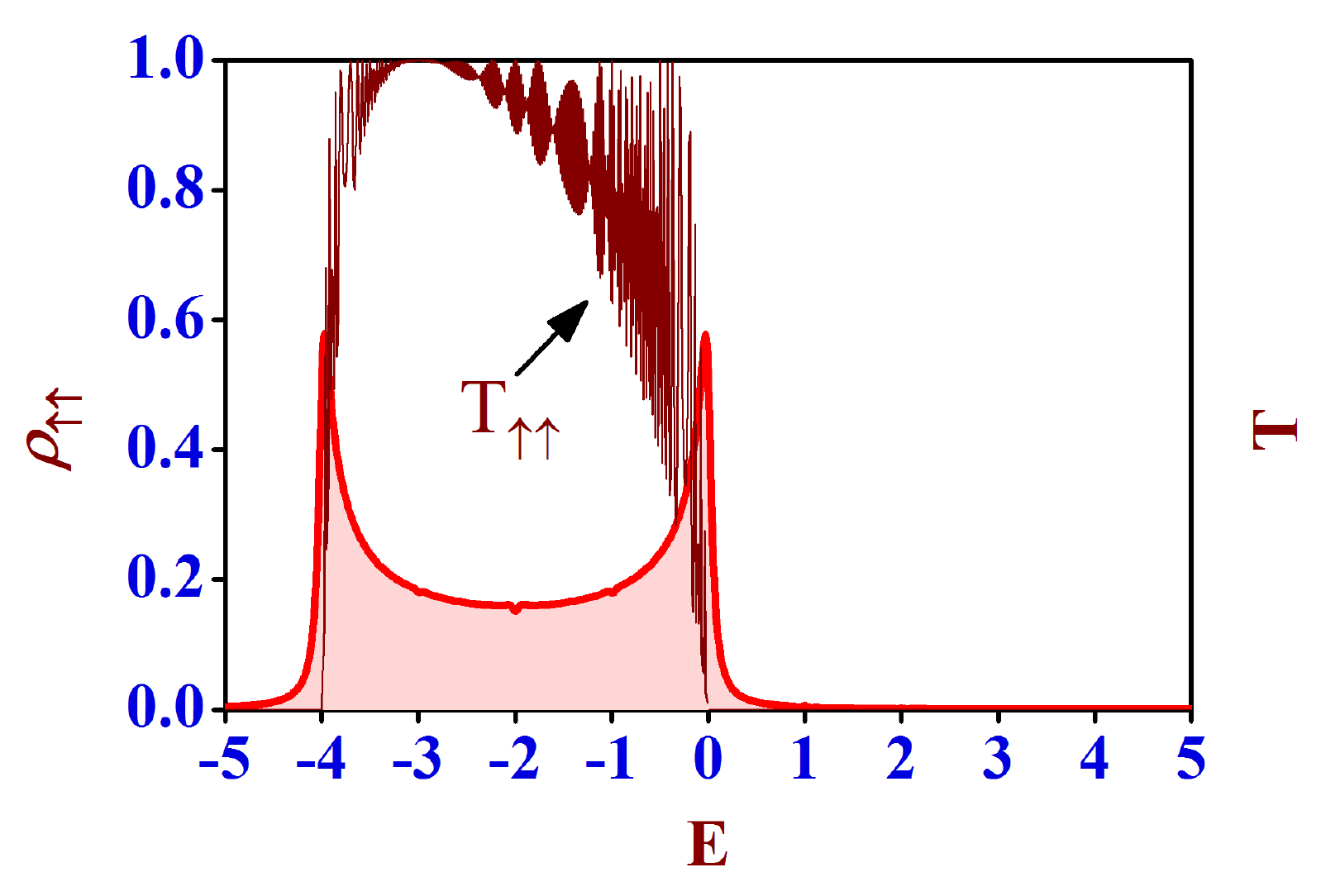}
(d)\includegraphics[clip,width=0.6\columnwidth, angle=0]{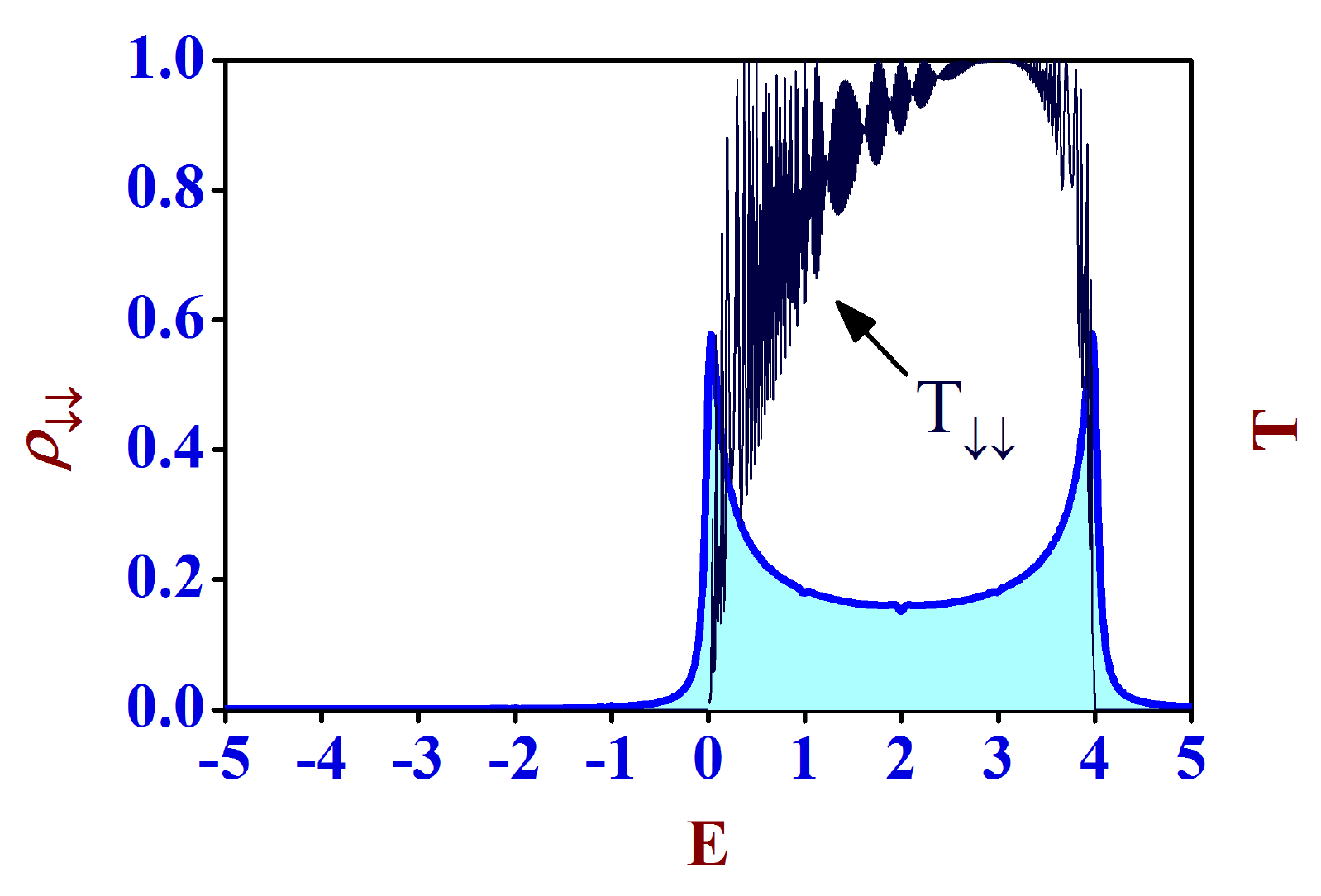}
\caption{(Color online) Plot of LDOS and transmission probabilities 
for spin-$1/2$ particles for 
spiral configurations of the magnetic moments in the magnetic chain 
system. The figures in the top panel [(a) and (b)] is for a variation 
of the angle $\theta_{n}=n\pi/L$, and the figures in the bottom 
panel [(c) and (d)] correspond to an angle variation $\theta_{n}=n\pi/(L/2)$ 
with $L=300$. We have set $h=2$ for all the magnetic sites in the chain.} 
\label{ldos-trans-half-spiral-L-Lby2}
\end{figure*}

In Fig.~\ref{ldos-trans-half-spiral-L-Lby2} we show the variation of 
the LDOS and corresponding transmission spectrum for a spin-$1/2$ spiral 
configuration. The periods of the spiral configurations are chosen to 
be $600$ and $300$ corresponding to 
Fig.~\ref{ldos-trans-half-spiral-L-Lby2}(a), (b) and (c), (d) respectively. 
We have set $h=2$ throughout and have taken $L=300$. The energy bands 
for the `up' and the `down' spin channels touch each other at this $h$ 
as in Figs.\ \ref{ldos-trans-half-theta0} and \ref{ldos-trans-half-thetaPiby4}. 
However, in contradistinction to these previous results, we now observe 
a \emph{spin-flipped transmission} in 
Fig.~\ref{ldos-trans-half-spiral-L-Lby2}(a), and (b). This can be 
understood if we recall that the `effective' on-site potentials for 
the `up' and the `down' spin electrons turn out to be 
$-h \cos(n\pi/L)$ and $h \cos(n \pi/L)$ respectively. The period of 
variation in $\theta_{n}$ is $2L$. Thus, for a system size equal to 
half the period, the incoming `up' particle traverses the potential 
landscape \emph{uphill}, as shown in Fig.\ \ref{spiralchain}(c). The 
transport of `up' spin electrons thus experiences resistance in 
traversing the system, and is eventually blocked over the energy 
range $-4 < E < 0$, though $\rho_{\uparrow\uparrow}$ is finite here.
On the other hand, the `down' spin electrons effectively move 
{\it downhill} on the potential landscape, and are transmitted 
in the same energy range. The complementary picture is visible 
in Fig.~\ref{ldos-trans-half-spiral-L-Lby2}(b). The specific 
choice of the polar angle $\theta_{n} = n\pi/L$ thus makes the 
$L$-atom long system a {\it spin flipper}.

The argument laid down here helps us understand the remaining two 
figures viz., Fig.~\ref{ldos-trans-half-spiral-L-Lby2} (c) and (d). 
We have now selected $\theta_{n} = n \pi/(L/2)$ and a system of 
length $L$ offers a full period. The substrate atom at the first 
site $n=0$ has its moment `up'. As the probe reaches the end, the 
substrate moment is back in the `up' orientation, favoring the `up' 
spin transport. The case is similar for the `down' spin states, but 
in the complementary range of the energy. We do not have a spin 
flipper anymore.
The results for higher spin $S$ are presented in the 
supplement~\cite{supplement}. There will be $m_{S}$ `spin channels', 
e.g., for $S=1$, the central channel, corresponding to the spin 
projection $m_{S} = 0$, turns out to be always transmitting, while the 
channels corresponding to $m_{S} = \pm 1$ exhibit spin-flipped 
transport for $\theta_{n} = n \pi/L$. The spin flipping character 
is lost if the system size includes a full period of variation in 
the polar angle $\theta_{n}$, just as it was for the spin-$1/2$ case.
We have also checked that a variation of the simple harmonic 
spiral variation presented here, using 
e.g. $\theta_{n} \propto \sqrt{n}$ or $n^2$, does not alter 
the overall spin-flipping presented here.
\section{Conclusion}
\label{conclu}
In conclusion, we have presented a simple one dimensional chain of magnetic 
atoms which can act as spin filter or spin flipper for particles with 
arbitrary spins. The central idea depends crucially on a mapping of the 
problem for a spin-$S$ particle into that of a $2S+1$ strand ladder network. 
The magnetic moment of the substrate atoms turns out to be equivalent to the 
`inter-strand' tunnel hopping integral, which needs to be tailored to open up 
gaps in the \textcolor{black}{DOS}. This is shown to lead to the desired spin 
filtering effect. The analysis has been carried out for a wide variety of 
spins, going well beyond the standard spin-$1/2$ case. Higher spins are 
obtainable in appropriate atomic gases. The present work presents a unified 
model which can, in principle, through an appropriate substrate engineering, 
filter out arbitrarily large spin components, treating fermions and bosons 
on the same footing.

\textcolor{black}{Talking about the experimental aspects of the work, the model 
system proposed by us can actually be realized nowadays experimentally 
as well. With the present day advanced technologies people can image and 
manipulate the spin direction of individual magnetic atoms~\cite{serrate} to grow 
nanomagnets where exchange-coupled atomic magnetic moments form an array. 
Such tailor-made nanomagnets posses a rich variety of magnetic properties 
and can be explored as constituents of nanospintronics 
technologies~\cite{alexander1}. So the 
model magnetic chains proposed by us are not far from reality. The concept of 
spin transport for higher spin states can be realized in atomic gases, and 
may lead to some exciting new generation spin-based devices. The idea of 
filtering out one of the spin components for a certain energy regime can 
have useful application in spin-based logic gates~\cite{alexander2}.} 
\begin{acknowledgements}
BP would like to thank DST, India and British Council for providing the 
financial support through a Newton-Bhabha Fellowship, and acknowledges the 
University of Warwick for their kind hospitality during his stay at Warwick. 
Special thanks must be given to all the members of DisQS group and 
Paramita Dutta for stimulating discussions during the course of this work.  
No new data sets were presented in this work as classified according to 
the UK open data policy.
\end{acknowledgements}
\appendix*
\section{Formulations to obtain the transport characteristics}
\subsection{Spin $1/2$ case}
We have used the transfer matrix method (TMM) to get the transmission 
probabilities for different spin channels. In this appendix we present 
a detailed formulation of that. 
We can easily recast Eqs.~\eqref{spineq1} and \eqref{spineq2} to have 
following matrix equation,
\begin{widetext}
\begin{equation}
\left ( \def\arraystretch{1.5} \begin{array}{c}
\psi_{n+1,\uparrow} \\
\psi_{n+1,\downarrow} \\
\psi_{n,\uparrow} \\
\psi_{n,\downarrow}
\end{array} \right )
=\underbrace{
\left ( \def\arraystretch{1.5} \begin{array}{cccc}
\dfrac{\left(E-\epsilon_{n,\uparrow}+h_{n}\cos\theta_{n} \right)}{t} & 
\dfrac{h_{n} \sin\theta_{n} e^{-i\phi_{n}}}{t} & -1 & 0 \\
\dfrac{h_{n} \sin\theta_{n} e^{i\phi_{n}}}{t} & 
\dfrac{\left(E-\epsilon_{n,\downarrow}-h_{n}\cos\theta_{n} \right)}{t} & 0 & -1\\
1 & 0 & 0 & 0\\
0 & 1 & 0 & 0
\end{array} \right )}_{\mbox{\boldmath $P_{n}$}}
\left ( \def\arraystretch{1.5} \begin{array}{c}
\psi_{n,\uparrow} \\
\psi_{n,\downarrow} \\
\psi_{n-1,\uparrow} \\
\psi_{n-1,\downarrow}
\end{array} \right ).
\label{tm}
\end{equation}
\end{widetext}
where ${\mbox{\boldmath $P_{n}$}}$ is the {\it transfer matrix} for the 
$n$-th site. 

We have a system (magnetic chain) with $N$ number of magnetic 
sites connected between two semi-infinite non-magnetic leads. 
So the matrix equation connecting the wave 
functions of the lead-system-lead bridge is given by,
\begin{equation}
\left ( \def\arraystretch{1.5} \begin{array}{c}
\psi_{N+2,\uparrow} \\
\psi_{N+2,\downarrow} \\
\psi_{N+1,\uparrow} \\
\psi_{N+1,\downarrow}
\end{array} \right )
=
\underbrace{{\mbox{\boldmath $M_{R}\cdot P\cdot M_{L}$}}}_{{\mbox{\boldmath $M$}}}
\left ( \def\arraystretch{1.5} \begin{array}{c}
\psi_{0,\uparrow} \\
\psi_{0,\downarrow} \\
\psi_{-1,\uparrow} \\
\psi_{-1,\downarrow}
\end{array} \right ).
\label{totaltm}
\end{equation}
where ${\mbox{\boldmath $M_{L}$}}$ is the transfer matrix for the left lead, 
${\mbox{\boldmath $M_{R}$}}$ is the transfer matrix for the right lead, 
${\mbox{\boldmath $P$}}$ $=$ $\prod_{n=N}^{1}{\mbox{\boldmath $P_{n}$}}$, and 
${\mbox{\boldmath $M$}}$ is the total transfer matrix for the lead-system-lead 
bridge.

\subsection{\it Evaluation of ${\mbox{\boldmath $M_{L}$}}$}
We have set $\epsilon_{L}=\epsilon_{0}$ for the all the sites in the left lead. 
The difference equation connecting the wave function amplitude of the $0$-th site 
with that of the $1$-th and $-1$-th sites is,
\begin{equation}
(E-\epsilon_{0})\psi_{0} = t_{LD}\psi_{1} + t_{L}\psi_{-1}.
\label{diffeq1}
\end{equation}
In the lead, according to the tight-binding model, we have
$\psi_{n} = Ae^{ikna}$ and $\psi_{0} = e^{i\gamma_{L}}\psi_{-1}$,
where $\gamma_{L}=ka$ and $E=\epsilon_{0}+2t_{L}\cos\gamma_{L}$.
Consequently we find 
$\psi_{1} = \left( t_{L} e^{i\gamma_{L}} / t_{LD} \right) \psi_{0}$.
In TMM form, this gives
\begin{equation}
\left ( \def\arraystretch{1.5} \begin{array}{c}
\psi_{1,\uparrow} \\
\psi_{1,\downarrow} \\
\psi_{0,\uparrow} \\
\psi_{0,\downarrow}
\end{array} \right )
=\underbrace{
\left ( \def\arraystretch{1.5} \begin{array}{cccc}
\dfrac{t_{L}}{t_{LD}}e^{i\gamma_{L}} & 0 & 0 & 0 \\
0 & \dfrac{t_{L}}{t_{LD}}e^{i\gamma_{L}} & 0 & 0 \\
0 & 0 & e^{i\gamma_{L}} & 0 \\
0 & 0 & 0 & e^{i\gamma_{L}}
\end{array} \right )}_{\mbox{\boldmath $M_{L}$}}
\left ( \def\arraystretch{1.5} \begin{array}{c}
\psi_{0,\uparrow} \\
\psi_{0,\downarrow} \\
\psi_{-1,\uparrow} \\
\psi_{-1,\downarrow}
\end{array} \right ).
\label{mL}
\end{equation}

\subsection{\it Evaluation of ${\mbox{\boldmath $M_{R}$}}$}
We set $\epsilon_{R}=\epsilon_{0}$ for the all the sites in the right lead. 
The difference equation connecting the wave function amplitude of the 
$(N+1)$-th site with that of the $(N+2)$-th and $N$-th sites is,
\begin{equation}
(E-\epsilon_{0})\psi_{N+1} = t_{R}\psi_{N+2} + t_{RD}\psi_{N},
\label{diffeq2}
\end{equation}
where
$\psi_{N+2} = Ae^{ik(N+2)a}$; $\psi_{N+2} = e^{i\gamma_{R}}\psi_{N+1}$,
and $\gamma_{R}=ka$ and $E=\epsilon_{0}+2t_{R}\cos\gamma_{R}$.
Consequently, 
$\psi_{N+1} = \left( t_{RD}  e^{i\gamma_{R}} / t_{R} \right) \psi_{N}$ and
\begin{equation}
\left ( \def\arraystretch{1.5} \begin{array}{c}
\psi_{N+2,\uparrow} \\
\psi_{N+2,\downarrow} \\
\psi_{N+1,\uparrow} \\
\psi_{N+1,\downarrow}
\end{array} \right )
=\underbrace{
\left ( \arraycolsep=0pt \def\arraystretch{1.5} \begin{array}{cccc}
e^{i\gamma_{R}} & 0 & 0 & 0 \\
0 & e^{i\gamma_{R}} & 0 & 0 \\
0 & 0 & \dfrac{t_{RD}}{t_{R}}e^{i\gamma_{R}} & 0 \\
0 & 0 & 0 & \dfrac{t_{RD}}{t_{R}}e^{i\gamma_{R}}
\end{array} \right )}_{\mbox{\boldmath $M_{R}$}}
\left ( \def\arraystretch{1.5} \begin{array}{c}
\psi_{N+1,\uparrow} \\
\psi_{N+1,\downarrow} \\
\psi_{N,\uparrow} \\
\psi_{N,\downarrow}
\end{array} \right ).
\label{mR}
\end{equation}

\subsubsection{An incoming spin up $(\uparrow)$}

If the incoming particle to the left electrode (lead) has a 
spin-up ($\uparrow$) projection then the wavefunction amplitudes 
in Eq.~\eqref{totaltm} can be written as,
$\psi_{-1,\uparrow} = e^{-i\gamma_{L}} + \mathcal{R}_{\uparrow \uparrow} e^{i\gamma_{L}}$, 
$\psi_{-1,\downarrow} = \mathcal{R}_{\uparrow \downarrow} e^{i\gamma_{L}}$, 
$\psi_{0,\uparrow} = 1 + \mathcal{R}_{\uparrow \uparrow}$, $\psi_{0,\downarrow} = \mathcal{R}_{\uparrow \downarrow}$,
$\psi_{N+2,\uparrow} = \mathcal{T}_{\uparrow \uparrow} e^{i(N+2)\gamma_{R}}$, 
$\psi_{N+2,\downarrow} = \mathcal{T}_{\uparrow \downarrow} e^{i(N+2)\gamma_{R}}$, 
$\psi_{N+1,\uparrow} = \mathcal{T}_{\uparrow \uparrow} e^{i(N+1)\gamma_{R}}$, 
$\psi_{N+1,\downarrow} = \mathcal{T}_{\uparrow \downarrow} e^{i(N+1)\gamma_{R}}$,
where $\mathcal{R}_{\uparrow \uparrow (\uparrow \downarrow)}$ and 
$\mathcal{T}_{\uparrow \uparrow (\uparrow \downarrow)}$ are the amplitudes 
of the reflected and transmitted electron wavefunctions with spin-up ($\uparrow$) 
projection, which remain in spin-up ($\uparrow$) state (or, flip to 
spin-down ($\downarrow$) state) after passing through the system. 
If we put the above values of the wavefunction amplitudes in 
Eq.~\eqref{totaltm} then we will have
\begin{equation}
\left ( \def\arraystretch{1.5} \begin{array}{c}
\mathcal{T}_{\uparrow \uparrow} e^{i(N+2)\gamma_{R}} \\
\mathcal{T}_{\uparrow \downarrow} e^{i(N+2)\gamma_{R}} \\
\mathcal{T}_{\uparrow \uparrow} e^{i(N+1)\gamma_{R}} \\
\mathcal{T}_{\uparrow \downarrow} e^{i(N+1)\gamma_{R}}
\end{array} \right )
=
{\mbox{\boldmath $M$}}\ 
\left ( \def\arraystretch{1.5} \begin{array}{c}
1 + \mathcal{R}_{\uparrow \uparrow} \\
\mathcal{R}_{\uparrow \downarrow} \\
e^{-i\gamma_{L}} + \mathcal{R}_{\uparrow \uparrow} e^{i\gamma_{L}} \\
\mathcal{R}_{\uparrow \downarrow} e^{i\gamma_{L}}
\end{array} \right ).
\label{totaltm-spinup}
\end{equation} 
We solve Eq.~\eqref{totaltm-spinup} for 
$\mathcal{T}_{\uparrow \uparrow}$ and $\mathcal{T}_{\uparrow \downarrow}$ and obtain the transmission probabilities as
\begin{equation}
T_{\uparrow \uparrow} = 
\dfrac{t_{R}\sin \gamma_{R}}{t_{L}\sin \gamma_{L}} 
\big|\mathcal{T}_{\uparrow \uparrow}\big|^{2}, \quad
T_{\uparrow \downarrow} = 
\dfrac{t_{R}\sin \gamma_{R}}{t_{L}\sin \gamma_{L}} 
\big|\mathcal{T}_{\uparrow \downarrow}\big|^{2}.
\label{tranUD}
\end{equation}
The total transmission probability for a spin-up ($\uparrow$) particle is 
given by,
\begin{equation}
T_{\uparrow} = T_{\uparrow \uparrow} + T_{\uparrow \downarrow}.
\label{tranU}
\end{equation}

\subsubsection{An incoming spin down $(\downarrow)$}

If the incoming particle to the left electrode (lead) has a 
spin down ($\downarrow$) projection then
$\psi_{-1,\uparrow} = \mathcal{R}_{\downarrow \uparrow} e^{i\gamma_{L}}$, 
$\psi_{-1,\downarrow} = e^{-i\gamma_{L}} + \mathcal{R}_{\downarrow \downarrow} e^{i\gamma_{L}}$, 
$\psi_{0,\uparrow} = \mathcal{R}_{\downarrow \uparrow}$, 
$\psi_{0,\downarrow} = 1 + \mathcal{R}_{\downarrow \downarrow}$, 
$\psi_{N+2,\uparrow} = \mathcal{T}_{\downarrow \uparrow} e^{i(N+2)\gamma_{R}}$, 
$\psi_{N+2,\downarrow} = \mathcal{T}_{\downarrow \downarrow} e^{i(N+2)\gamma_{R}}$, 
$\psi_{N+1,\uparrow} = \mathcal{T}_{\downarrow \uparrow} e^{i(N+1)\gamma_{R}}$, 
$\psi_{N+1,\downarrow} = \mathcal{T}_{\downarrow \downarrow} e^{i(N+1)\gamma_{R}}$, 
where $\mathcal{R}_{\downarrow \downarrow (\downarrow \uparrow)}$ and 
$\mathcal{T}_{\downarrow \downarrow (\downarrow \uparrow)}$ are the amplitudes 
of the reflected and transmitted electron wavefunctions with spin-down ($\downarrow$) 
projection, which remain in spin-down ($\downarrow$) state (or, flip to 
spin-up ($\uparrow$) state) after passing through the system. 
As before, we find
\begin{equation}
\left ( \def\arraystretch{1.5} \begin{array}{c}
\mathcal{T}_{\downarrow \uparrow} e^{i(N+2)\gamma_{R}} \\
\mathcal{T}_{\downarrow \downarrow} e^{i(N+2)\gamma_{R}} \\
\mathcal{T}_{\downarrow \uparrow} e^{i(N+1)\gamma_{R}} \\
\mathcal{T}_{\downarrow \downarrow} e^{i(N+1)\gamma_{R}}
\end{array} \right )
=
{\mbox{\boldmath $M$}}\ 
\left ( \def\arraystretch{1.5} \begin{array}{c}
\mathcal{R}_{\downarrow \uparrow} \\
1 + \mathcal{R}_{\downarrow \downarrow} \\
\mathcal{R}_{\downarrow \uparrow} e^{i\gamma_{L}} \\
e^{-i\gamma_{L}} + \mathcal{R}_{\downarrow \downarrow} e^{i\gamma_{L}}
\end{array} \right ),
\label{totaltm-spindown}
\end{equation} 
and solve Eq.~\eqref{totaltm-spindown} for 
$\mathcal{T}_{\downarrow \downarrow}$ and $\mathcal{T}_{\downarrow \uparrow}$. 
The transmission probabilities are now 
\begin{equation}
T_{\downarrow \downarrow} = 
\dfrac{t_{R}\sin \gamma_{R}}{t_{L}\sin \gamma_{L}} 
\big|\mathcal{T}_{\downarrow \downarrow}\big|^{2}, \quad
T_{\downarrow \uparrow} = 
\dfrac{t_{R}\sin \gamma_{R}}{t_{L}\sin \gamma_{L}} 
\big|\mathcal{T}_{\downarrow \uparrow}\big|^{2},
\label{tranDU}
\end{equation}
with total transmission probability for a spin-down ($\downarrow$) particle 
\begin{equation}
T_{\downarrow} = T_{\downarrow \downarrow} + T_{\downarrow \uparrow}.
\label{tranD}
\end{equation}
\subsection{An example for higher spin cases: spin $1$}
Proceeding in the same way as in the spin-$1/2$ case, we obtain the 
following forms of the transfer matrices for the left and 
right leads as,
$\bm{M_L}= e^{i\gamma_{L}} \mathrm{diag}\left(
t_{L}/t_{LD}, t_{L}/t_{LD}, t_{L}/t_{LD}, 1, 1, 1 \right)$
and
$\bm{M_R}= e^{i\gamma_{R}} \mathrm{diag}\left(
1, 1, 1, t_{R}/t_{RD}, t_{R}/t_{RD}, t_{R}/t_{RD}\right)$.
The transfer matrix for spin-$1$ particles at the 
$n$-th site reads as,
\begin{widetext}
\begin{equation}
{\mbox{\boldmath $P_{n}$}} = 
\left ( \def\arraystretch{1.5} \begin{array}{cccccc}
\dfrac{\left(E-\epsilon_{n,1}+h_{n}\cos\theta_{n} \right)}{t} & 
\dfrac{h_{n} \sin\theta_{n} e^{-i\phi_{n}}}{\sqrt{2}t} & 
0 & -1 & 0 & 0\\
\dfrac{h_{n} \sin\theta_{n} e^{i\phi_{n}}}{\sqrt{2}t} & 
\dfrac{\left(E-\epsilon_{n,0}\right)}{t} & 
\dfrac{h_{n} \sin\theta_{n} e^{-i\phi_{n}}}{\sqrt{2}t} & 
0 & -1 & 0\\
0 & \dfrac{h_{n} \sin\theta_{n} e^{i\phi_{n}}}{\sqrt{2}t} & 
\dfrac{\left(E-\epsilon_{n,-1}-h_{n}\cos\theta_{n} \right)}{t} & 
0 & 0 & -1\\
1 & 0 & 0 & 0 & 0 & 0\\
0 & 1 & 0 & 0 & 0 & 0\\
0 & 0 & 1 & 0 & 0 & 0
\end{array} \right ).
\label{tmspin1}
\end{equation}
\end{widetext}
Similar to the spin-$1/2$ particles, we obtain the transmission probabilities 
for the spin-$1$ particles as,
\begin{equation}
T_{\sigma \sigma^{\prime}} = 
\dfrac{t_{R}\sin \gamma_{R}}{t_{L}\sin \gamma_{L}} 
\big|\mathcal{T}_{\sigma \sigma^{\prime}}\big|^{2},
\label{alltran}
\end{equation}
where $\sigma, \sigma^{\prime} = 1, 0, -1$, and the total transmission 
probabilities for spin $1$, $0$, and $-1$ components, respectively, are given by,
\begin{subequations}
\label{transmission1}
\begin{eqnarray}
& T_{1} = T_{1,1} + T_{1,0} + T_{1,-1}, \\ 
\label{tran1}
& T_{0} = T_{0,1} + T_{0,0} + T_{0,-1}, \\
\label{tran0}
& T_{-1} = T_{-1,1} + T_{-1,0} + T_{-1,-1}.
\label{tranM1}
\end{eqnarray}
\end{subequations}
%
Clearly, this scheme can be carried forward to obtain the transport 
characteristics for particles with any arbitrary spin. For spin 
$S=1/2$ in~\eqref{tm} and for $S=1$ in~\eqref{tmspin1}, the on-site 
magnetic strength coefficients $\propto \cos\theta_n$ are 
$1$, $-1$ and $1$, $0$, $-1$, respectively. For $S=3/2$, we have 
$1,1/3,-1/3,-1$, for $S=2$ they are $1, 1/2,0,-1/2,-1$ and 
for $S=5/2$, we have $1,3/5,1/5,-1/5,-3/5,-1$.
The proportionality coefficients for the hopping 
$\propto\sin\theta_{n}$ are $1$ for $S=1/2$, and $1/\sqrt{2}, 1/\sqrt{2}$ 
for $S=1$ in \eqref{tm}  and \eqref{tmspin1}, respectively. 
For $S=3/2$, we find $1/\sqrt{3}, 2/3, 1/\sqrt{3}$, for $S=2$ 
they are $1/2, \sqrt{6}/4, \sqrt{6}/4, 1/2$ and, last, 
for $S=5/2$ we have $1/\sqrt{5}, \sqrt{8}/5, 3/5, \sqrt{8}/5, 1/\sqrt{5}$. 
These coefficients, for the $\cos\theta_{n}$ terms, 
are used in Fig.~\ref{spiralchain}(c).


\end{document}